\documentclass[conference]{IEEEtran}
\usepackage[table,xcdraw]{xcolor}
\usepackage{url}
\usepackage{hyperref}
\usepackage{color, soul}
\usepackage{graphicx}
\usepackage{rotating}
\usepackage{booktabs}
\usepackage{pifont}
\usepackage[center]{caption}
\usepackage{todonotes}
	
\usepackage{rotating}
\usepackage[T1]{fontenc}
\usepackage{circledsteps}
\usepackage[normalem]{ulem}
\useunder{\uline}{\ul}{}
\usepackage[compress]{cite}
\usepackage{enumitem}
\usepackage{fontawesome}
\usepackage{bbding}
\usepackage{enumitem}

\newcolumntype{M}[1]{>{\arraybackslash}m{#1}}
\usepackage{multirow}

\newcommand{\eln}[1]{{\color{black} #1}}
\newcommand{\halla}[1]{{\color{black} #1}}
\newcommand{\gmit}[1]{{\color{black} #1}}

\begin{document}

\title{Systematically Assessing the Security Risks of AI/ML-enabled Connected Healthcare Systems}

%
%
%



\author{\IEEEauthorblockN{Mohammed Elnawawy\textsuperscript{$\dagger$}, Mohammadreza Hallajiyan\textsuperscript{$\dagger$}, Gargi Mitra\textsuperscript{$\dagger$}, Shahrear Iqbal\textsuperscript{$\ast$}, Karthik Pattabiraman\textsuperscript{$\dagger$}}

\IEEEauthorblockA{
\textsuperscript{$\dagger$}University of British Columbia, \textsuperscript{$\ast$}National Research Council Canada\\
Email: \{mnawawy, hallaj, gargi\}@ece.ubc.ca, shahrear.iqbal@nrc-cnrc.gc.ca, karthikp@ece.ubc.ca}

}

\IEEEpubid{0000--0000/00\$00.00~\copyright~2021 IEEE}

\maketitle

\begin{abstract}
The adoption of machine-learning-enabled systems in the healthcare domain is on the rise. While the use of ML in healthcare has several benefits, it also expands the threat surface of medical systems. We show that the use of ML in medical systems, particularly connected systems that involve interfacing the ML engine with multiple peripheral devices, has security risks that might cause life-threatening damage to a patient's health in case of adversarial interventions. These new risks arise due to security vulnerabilities in the peripheral devices and communication channels. We present a case study where we demonstrate an attack on an ML-enabled blood glucose monitoring system by introducing adversarial data points during inference. We show that an adversary can achieve this by exploiting a known vulnerability in the Bluetooth communication channel connecting the glucose meter with the ML-enabled app. 
We further show that state-of-the-art risk assessment techniques are not adequate for identifying and assessing these new risks. 
Our study highlights the need for novel risk analysis methods for analyzing the security of AI-enabled connected health devices.
\end{abstract}

\begin{IEEEkeywords}
Machine learning, FDA, medical system security, risk analysis, multi-vendor systems.
\end{IEEEkeywords}

\section{Introduction}\label{sec:intro}

The use of Artificial Intelligence (AI), especially Machine Learning (ML) techniques, is becoming increasingly popular in the medical field. As of October 2022, the U.S. Food and Drug Administration (FDA) has approved $521$ ML-enabled devices across $15$ different medical disciplines (e.g., Cardiology, Ophthalmology, and Gastroenterology)~\cite{fdaml}. However, 
the use of ML has expanded the threat surface of medical systems~\cite{chen2020ecgadv,albattah2023detection,lal2021adversarial,9313421,joel2021adversarial,bortsova2021adversarial,menon2021covid,vargas2020understanding,levy2022personalized,chen2022adversarial,yu2023perturbing,nielsen2022investigating,mangaokar2020jekyll,hu2022adversarial,ma2021understanding} 
making them more vulnerable to cyberattacks.

ML-enabled medical devices are used for performing critical activities such as remote patient monitoring, controlling surgical equipment, automatic drug administration, and preliminary/advanced disease diagnosis
-- tasks that require 
high accuracy and reliability~\cite{fdaml}. 
If an adversary compromises such a device, it can force the ML engine to make
incorrect predictions or decisions, which can have catastrophic consequences, such as wrong treatment leading to health complications. 

An adversary can force an ML engine to generate incorrect predictions or decisions by injecting carefully crafted malicious data points either during training or inference. Preventing such attacks in ML-enabled medical devices is challenging.
These ML-enabled devices are typically interconnected with other peripheral sensor devices that collect physiological data of patients, which are then processed by the ML engine. Therefore, it is not enough to secure the ML-enabled device, since adversaries can exploit vulnerabilities in the peripheral devices to inject malicious data points in the ML engine. 

To protect the end-to-end system, one must systematically identify and assess the 
security risks~\footnote{We define risk as the probability of a security vulnerability getting exploited, and its potential impact or loss.} of the overall system due to vulnerabilities in peripheral devices. \emph{To the best of our knowledge, there is no systematic technique for identifying and assessing the end-to-end risks of ML-enabled medical systems.}

\IEEEpubidadjcol

Identification of risks in ML-enabled connected medical systems has two challenges. \textit{First,} at deployment, the ML-enabled device is interfaced with several other peripheral devices, each of which may be manufactured by a different company. For instance, a user of the ML-enabled blood glucose monitoring app Dreamed Advisor Pro~\cite{dreamed}, needs to install the app on a smartphone, and then connect to it a smartwatch, a glucose meter, and an insulin pump, all of which would be manufactured by different companies, and may have their own security vulnerabilities. \textit{Second,} each app user may use peripheral devices from different sets of manufacturers, leading to diverse vulnerabilities among different users of the same app. For instance, one user of the app might use a vulnerable smartphone, while another user might use a vulnerable glucose meter. 

Furthermore, it is also challenging to assess the \emph{severity} of these risks. This is because the severity of a risk posed by a vulnerable peripheral device might differ when assessed in the context of the individual device (in-silo assessment) versus when assessed in the context of the entire system. For instance, consider a user who connects the Dreamed Advisor Pro app to a glucose meter with a write-access vulnerability, and a smartphone with a read-access vulnerability. When assessed separately, the glucose meter would have a higher perceived risk than the smartphone. However, for an adversary who wants to inject adversarial glucose meter readings into the app, being able to read data from the smartphone (e.g., meal timings, latest insulin dose, carbohydrates taken) might be useful for crafting malicious data points that adhere to physiological constraints. Adhering to the physiological constraints is important for the adversary to get the malicious data points accepted as valid inputs by the ML engine. Therefore, we need to holistically consider the risks from the interplay of vulnerabilities in peripheral devices.  

In this paper, we perform a systematic analysis to highlight the security risks of end-to-end ML-enabled connected medical systems. Our analysis consists of three steps. 
\textit{First}, we conduct a systematic exploration of the FDA-approved ML-enabled medical devices to understand the ML techniques that they use, and the damage that can be caused to a patient if the ML technique mispredicts their case. \textit{Second}, we conduct an extensive review to \textit{identify} possible ways in which adversaries can inject malicious data points into an ML-enabled medical device at deployment. This involves a cross-domain analysis, where we map known attacks on ML algorithms with known vulnerabilities in peripheral devices that would make the attacks practical. 
\textit{Finally}, we perform a critical evaluation of state-of-the-art risk assessment frameworks used by the ML-enabled medical device manufacturing companies today. We identify the loopholes in these risk assessment strategies that might make manufacturers miss the risks arising due to vulnerabilities in connected peripheral devices. 

\noindent\textbf{Contributions. }The main contributions of this paper are:
\begin{enumerate}[leftmargin=*]
    \item We perform a systematic cross-domain security analysis of commercial ML-enabled medical devices approved by the FDA (Section \ref{sec:attacks}), to highlight the security risks of connected health devices.
    \item We then perform a case study on a realistic ML-enabled blood glucose management system (BGMS) (Section \ref{sec:case-study}) to demonstrate an attack on the system where the adversary compromises a communication link in the system. 
    \item Finally, we perform an evaluation of state-of-the-art risk assessment techniques (Section \ref{sec:existingframeworks}). We find that they are inadequate in identifying and analyzing the severity of security risks in ML-enabled medical systems, particularly the risks posed to the ML engine by vulnerable peripheral devices. We also highlight directions for improvement. 

\end{enumerate}

\section{Motivation and Background}\label{sec:motivation}
We highlight security risks in AI/ML-enabled medical systems due to vulnerabilities in their connected peripheral components, using the example of the BGMS. Following this, we demonstrate the generalizability of identified risks to any connected ML-enabled medical system.


\subsection{Blood Glucose Management: Background}
Diabetes is a chronic health condition that hinders the body's natural insulin production capability, leading to elevated blood glucose levels. It has detrimental effects on a patient's health, and sudden spikes or drops in blood glucose can be life-threatening. 
Blood glucose levels can be divided into three ranges, hypoglycemic ($<$ 70-80 mg/dL), normal (80 mg/dl - 125 mg/dl), and hyperglycemic ($>$ 125 mg/dL while fasting and $>$ 180 mg/dL two hours postprandial)\cite{misc_diabetes_34, Mouri_Badireddy_2023}. A consistently hyperglycemic patient (i.e., diabetic) requires insulin injections to normalize their glucose levels. In contrast, a hypoglycemic patient does not need insulin injections.

 

BGMS apps help diabetic patients monitor their blood glucose levels and administer insulin bolus whenever the glucose levels begin to rise at an abnormal rate. 
The patient can either manually inject the insulin, or use an automated insulin pump connected to and controlled by the BGMS app. It is crucial to calculate the insulin bolus dose accurately — an overdose can lead to a sharp drop in blood glucose, while an insufficient dose may not bring it down to the normal range.



\subsection{An ML-enabled Blood Glucose Management System}\label{sec:motiv-bgms}
We consider a commercial FDA-approved ML-enabled BGMS app, the Dreamed Advisor Pro~\cite{dreamed}. 
This app assists diabetic patients in maintaining normal blood glucose levels by periodically recommending insulin bolus doses, personalized meal plans, and physical activities. 
However, since the specific ML technique used by the Dreamed Advisor Pro app is not publicly disclosed, we instead use a well-known, public ML-based blood glucose prediction technique~\cite{rubin2020deep}. 

BGMS apps suggest insulin doses based on the patient's predicted blood glucose level in the near future (next $30$ or $60$ mins)~\cite{rubin2020deep}. This prediction is done by an ML engine running at the back-end of the BGMS app, using recent physiological values of the patient, such as blood glucose measurements, insulin doses taken, meal timings, carbohydrate intake, etc. These values are either entered manually into the app by the patient or read from other peripheral sensor devices (e.g., continuous glucose monitors and smartwatches) connected to the patient's body. Besides sensor devices, the app can interface with actuator devices (e.g., insulin pumps) to execute actions suggested by the ML-enabled app. Together, the ML-enabled app and connected peripheral devices form the BGMS. 

\begin{figure}[t]
    \centering
    \includegraphics[scale=0.15]{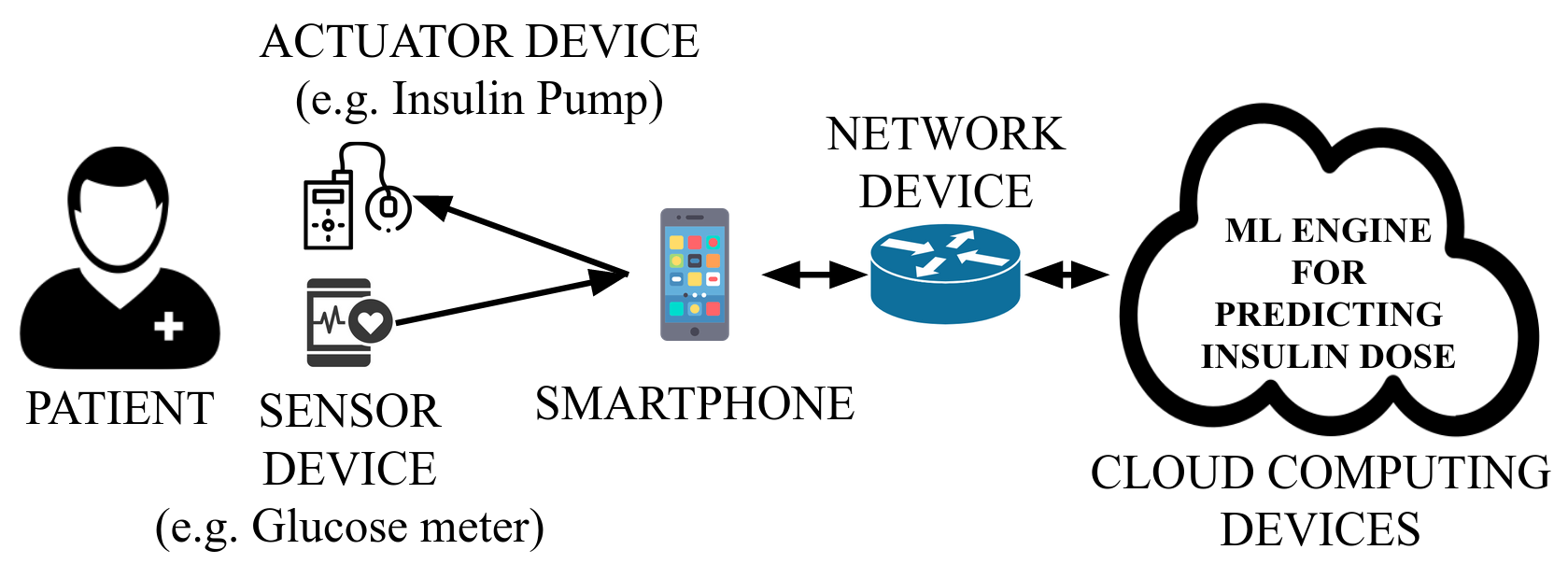}
    \caption[justification=centering]{A blood glucose management system that uses ML on the cloud, and interfaces with multiple peripheral devices. }
    \label{fig:bgms}
\end{figure}

Figure~\ref{fig:bgms} shows an end-to-end schematic representation of an ML-enabled BGMS. The glucose monitor records the blood glucose levels of the patient at regular time intervals and transmits them to the app over a Bluetooth communication channel. The app uses these values for data visualization and also sends them to a cloud server over the Internet for storage and processing by the AI/ML engine. Finally, the predicted insulin dose is either displayed on the app or sent to the automated insulin pump attached to the patient's body. 

\noindent\textbf{Security Risks. } We consider a scenario where an adversary intends to cause harm to a targeted user of the BGMS app by forcing the BGMS to inject a high dose of insulin into the user's body when it is not supposed to, or vice versa. We further assume that the insulin pump used by the target is secure against known vulnerabilities~\cite{rasmussen2022blurtooth} and that the ML engine runs on a secure cloud server. Under such circumstances, the adversary can still force the ML engine to mispredict the insulin dose by injecting carefully crafted adversarial data points into it via the peripheral devices. For instance, the adversary can modify the blood glucose values sent to the ML engine~\cite{levy2022personalized}, 
by exploiting vulnerabilities in the third-party glucose monitor interfaced with the BGMS app.

The Dreamed Advisor Pro app is compatible with all non-continuous glucose meters with regulatory approval and nine different models of continuous glucose monitors manufactured by different companies~\cite{Dreamedperipherals}. Commercial glucose meters have been known to have firmware vulnerabilities~\cite{dupont2020matter} that would allow an adversary to change the blood glucose level readings that are sent by the glucose meter to the BGMS app. Alternatively, an adversary could also exploit known vulnerabilities in the Bluetooth communication protocol~\cite{rasmussen2022blurtooth} to launch a man-in-the-middle attack and change the blood glucose level readings. Furthermore, these glucose monitors are also susceptible to physical attacks that can be carried out with electromagnetic radiation~\cite{mortazavi2014electromagnetic}. Additionally, vulnerabilities in devices and communication links in the end-to-end data processing pipeline, such as mobile devices and routers, 
can be exploited to manipulate blood glucose measurements.

\subsection{Unique Security Risks in AI/ML-enabled Medical Systems} 
The security risks discussed in the context of the BGMS app apply to any ML-enabled medical device connected to multiple peripheral devices (often manufactured by other companies) at deployment. A security breach in any of these peripheral devices could enable an adversary to manipulate data sent to the ML engine, resulting in mispredictions regarding the patient's condition. These mispredictions pose a direct threat to the patient's health.

Manufacturers of ML-enabled devices face two challenges in anticipating and assessing the aforementioned security risks during design and manufacturing. \textit{First,} these devices are built to be compatible with a diverse range of peripheral devices for the operational convenience of the consumers. This makes it difficult for the manufacturer to predict what peripheral devices the consumer would connect with the ML-enabled device at deployment, and what vulnerabilities those devices might have. 
The interplay of different vulnerabilities would enable an adversary to perform different types of attacks on the ML engine, which makes it challenging to analyze the risk via the in-silo testing performed today. 
\textit{Second,} while ML-enabled connected systems are used in many domains such as smart homes and industrial control systems, performing risk analysis of medical systems is more difficult as physiological data is much more complex and varies widely across individuals~\cite{hulsen2019big}. 
Consequently, the impact of manipulating physiological data might be different for different patients. Manufacturers typically prioritize accuracy and failures in non-adversarial scenarios, but a comprehensive end-to-end security risk analysis should consider the impact of adversarial inputs on different patients and the feasibility of attacks.
 


For instance, in the BGMS attack in Section \ref{sec:motiv-bgms}, the adversary attempts to generate inaccurate insulin dosage predictions by introducing adversarial inputs to the ML engine. 
Most known attacks on ML engines~\cite{chen2020ecgadv,albattah2023detection,lal2021adversarial,9313421,joel2021adversarial,bortsova2021adversarial,menon2021covid,vargas2020understanding,levy2022personalized,chen2022adversarial,yu2023perturbing,nielsen2022investigating,mangaokar2020jekyll,hu2022adversarial,ma2021understanding} require the adversary to observe and manipulate at least a subset of input sensor values to alter the predicted insulin dose. The adversary's efficiency in crafting adversarial inputs increases with greater observability into different sensor values. In Section \ref{sec:motiv-bgms}, we assumed the adversary could only manipulate glucose level readings through vulnerabilities in the glucose meter or the Bluetooth link. However, if the adversary can observe commands sent to the insulin pump, they could craft adversarial glucose meter readings to yield a higher success rate with equal or fewer perturbations. While unauthorized read access to an individual insulin pump poses a low-level risk, in the context of the entire BGMS, it becomes a high-level risk.

\section{Attacks on ML-Based Systems and Their Relevance in the Healthcare Domain}\label{sec:attacks}
Motivated by the BGMS example in Section \ref{sec:motivation}, 
we systematically investigate a subset of FDA-approved AI/ML-enabled medical devices/software~\cite {fdaml} to identify potential security risks at deployment. 
We perform this investigation in two steps. \textit{First,} we identify the ML techniques used by each of the devices/software systems. We survey existing work in the domain of AI/ML security to understand what types of attacks may target these techniques (Section \ref{sec:knownattacks}). \textit{Next,} for each of these devices/software, we examine the practicality of the attack scenarios identified in the previous step ( S\ref{sec:vuldevices}).   

\noindent\textbf{Selecting medical systems for our investigation.}
As of December 2022, the U.S. FDA has approved $521$ ML-enabled medical devices~\footnote{As per the terminology used by the FDA website, the term `device’ refers to both physical devices as well as software solutions. } across $15$ different physiological panels. However, there is no automated risk analysis technique today, and analyzing all the $521$ devices manually would be arduous and time-consuming. Therefore, 
we selected a subset of these devices for manual analysis. We used the following selection criteria to ensure a fair representation of the set of devices.
\begin{enumerate}[leftmargin=*]
    \item We select at least one device from each physiological panel to study if risks due to vulnerabilities in peripheral devices are common across all medical domains;
    \item Within the same physiological panel, we select devices that perform different types of diagnosis or treatments, to ensure coverage across different medical activities.
    \item We select an equal number of two types of ML-enabled devices - software that can be installed on the consumer's pre-existing device, and software that is sold bundled with proprietary hardware. This would help us understand if one of these is more secure than the other;
    \item We select devices that are used in hospitals and clinics under medical supervision, as well as devices that are used by patients at home without medical supervision. This would help us understand if the environment in which the medical device is deployed affects its security.
\end{enumerate}

Additionally, to ensure that sufficient information is available for each selected device, e.g., the ML algorithm used, the type of data processed, etc,  we  select $20$ different devices across $13$ of the $15$ physiological panels, as shown in Table~\ref{tab:knownmlattacks}. Unfortunately, due to insufficient information, we could not select any device from the Dental and Hematology panels. 


\subsection{Known Attacks on ML Algorithms Used by FDA-Approved Medical Devices}\label{sec:knownattacks}
Table~\ref{tab:knownmlattacks} presents our study of the ML algorithms used by the devices we selected for our evaluation. Our goal is to understand if there are known attacks against these ML algorithms that can be used by adversaries to make these ML engines mispredict the outcome. We also examine the types of tasks for which these ML algorithms are used, and the worst-case consequences of misprediction by the ML engine. 

\noindent\textbf{Survey Process. } We performed the following steps to identify known attacks on the ML algorithms used by the devices. 
    
\gmit{\noindent\textit{Step 1. Identifying the ML algorithm and input features used by the device:} We analyze device information available in the Premarket Notification summaries submitted by manufacturers to the FDA during the approval process. These summaries are available on the FDA website~\cite{fdaml}, and contain crucial details such as the ML algorithm and the type of data processed. However, for some devices, like GI Genius, the summaries lack specific information about the ML algorithm. In such cases, we explore the manufacturer's website and, if even that is inadequate, we estimate the ML algorithm based on the device's task and processed inputs. We look for known ML algorithms that perform the same task using similar input features. For example, for GI Genius, we found a relevant paper~\cite{georgakopoulos2016weakly} that performs gastrointestinal lesion detection (the same task performed by GI Genius) with a high accuracy using Convolutional Neural Networks (CNN).


\noindent\textit{Step 2. Identifying known attacks on the ML algorithm: }We search for known attacks in the literature that target the ML algorithms identified in Step 1. We focus on attacks described in research papers published in both conferences and journals. The discovery of such attacks does not definitively establish the vulnerability of the ML engine in the device under consideration. Rather, it identifies potential risks, emphasizing the need for systematic risk identification and mitigation.

\noindent\textit{Step 3. Estimating worst-case impact of mispredictions:} We estimate the worst-case impact of mispredictions by these devices from our understanding of the device functionality and description provided by the manufacturer, either in the device summary or on their website. We deem the misprediction to be potentially fatal if the device is used for the treatment or diagnosis of a patient in  medical emergencies, and a medical expert would not have enough time to assess the correctness of the device output. For instance, the NuVasive Pulse System is used by surgeons during spinal surgeries for continuously monitoring the neurophysiological status of the patient, and a misprediction by the device would be potentially fatal. 
}

\noindent\textbf{Insights. }We obtained the following insights from the information that we gathered using the aforementioned process. 
\begin{enumerate}[leftmargin=*]
    \item We observed that the majority of the ML algorithms are vulnerable to inference-time attacks, with a few susceptible to training-time attacks. Both of these pose major health risks for patients. However, executing inference-time attacks is comparatively easier for adversaries as they demand fewer adversarial inputs than training-time attacks. Most of the devices prone to training-time attacks are deployed in hospitals or diagnostic centers, where a shared set of peripheral devices is used to collect data from multiple patients. If adversaries successfully compromise these peripheral devices over an extended period, they can manipulate sufficient patient data to poison the training dataset~\footnote{Many systems undergo periodic re-training on recent physiological data.}.
    This would affect a large number of patients.
    Examples include the Deep Learning Image Recognition Software, and the Oxehealth Vital Signs monitor.
    \item Even when the devices are operated by medical practitioners, detecting a misprediction might be challenging for two reasons. \textit{First,} physiological data exhibit significant variance even among patients with the same medical condition, owing to diverse underlying health conditions and demographic factors~\cite{issa2007personalized}. 
    \textit{Second,} many devices are used for infrequently performed diagnoses/medical procedures, or are used only for medical emergencies. Under such circumstances, the lack of the particular patient's historical physiological information makes it challenging for the medical practitioner to detect an anomaly. Examples are Cardiologs ECG Analysis Platform, GI Genius, ABMD Software, and NuVasive Pulse System.
    \item Some devices (e.g., the One Drop Blood Glucose Monitoring System), are used by patients at home without continuous medical supervision. Detecting a misprediction from such devices would be much more challenging than devices that are directly operated by medical practitioners.
    \item Many of these devices are used in clinics and hospitals for disease diagnosis, treatment, and patient monitoring. However, a few are used by patients at home. 
    Hospitals and clinics would typically have  a higher security budget than individual patients at home, and hence have better  security. Consequently, designing a one-size-fits-all security solution for ML-enabled medical systems is challenging. Therefore, while designing security solutions for medical devices, the implementation effort and cost should be considered. 
    \item Some of the ML-enabled softwares are sold bundled with proprietary hardware (i.e., software-in-medical-device), while some can be installed by the user on any general-purpose computer (i.e., software-as-medical-device). The latter have a broader threat surface due to diverse combinations of hardware, software, and Operating System (OS) vulnerabilities across various general-purpose computer models, making the assessment of risk severity challenging.
      
\end{enumerate}


\subsection{Analyzing the Functionality and Vulnerability Landscape of FDA-approved (AI/ML)-Enabled Medical Devices}\label{sec:vuldevices}


\begin{table*}[hbt!]
\footnotesize
\setlength\tabcolsep{1pt}
\begin{tabular}{|c|c|c|c|c|c|c|c|c|}
\hline
\rowcolor[HTML]{EFEFEF} 
\textbf{\begin{tabular}[c]{@{}c@{}}Sl.\\ No.\end{tabular}} &
  \textbf{\begin{tabular}[c]{@{}c@{}}Device \\ Name~\cite{fdaml}\end{tabular}} &
  \textbf{\begin{tabular}[c]{@{}c@{}}Physiological\\ Panel\end{tabular}} &
  \textbf{\begin{tabular}[c]{@{}c@{}}Device\\ Functionality\end{tabular}} &
  \textbf{User} &
  \textbf{\begin{tabular}[c]{@{}c@{}}Type of \\ ML algo\\ used\end{tabular}} &
  \textbf{\begin{tabular}[c]{@{}c@{}}Type of\\ data\\ processed\end{tabular}} &
  \textbf{\begin{tabular}[c]{@{}c@{}}Known\\ attacks\\ (Attack type)\end{tabular}} &
  \textbf{\begin{tabular}[c]{@{}c@{}}Potential\\ impact of\\ misprediction\end{tabular}} \\ \hline
1 &
  \begin{tabular}[c]{@{}c@{}}CardioLogs \\ ECG Analysis\\ Platform$^{\dagger}$\end{tabular} &
  Cardiovascular &
  \begin{tabular}[c]{@{}c@{}}Cardiac \\ arrhythmia\\ detector\end{tabular} &
  \begin{tabular}[c]{@{}c@{}}Medical\\ practitioners\end{tabular} &
  \begin{tabular}[c]{@{}c@{}}Deep Neural Network\\ (DNN)\end{tabular} &
  Image &
  Chen et al.~\cite{chen2020ecgadv} \CircledTop{I} &
  \begin{tabular}[c]{@{}c@{}}Wrong \\ treatment\\ (Fatal)\end{tabular} \\ \hline
2 &
  \begin{tabular}[c]{@{}c@{}}Oxehealth \\ Vital Signs$^{\dagger}$\end{tabular} &
  Cardiovascular &
  \begin{tabular}[c]{@{}c@{}}Camera-based\\ monitor for heart,\\ pulse, and\\ respiratory rate\end{tabular} &
  \begin{tabular}[c]{@{}c@{}}Medical\\ practitioners\end{tabular} &
  \begin{tabular}[c]{@{}c@{}}Hybrid \\ convolutional Long\\  short term memory\\  networks (LSTM)\end{tabular} &
  Video &
  \begin{tabular}[c]{@{}c@{}}Albattah \\ et al.~\cite{albattah2023detection} \CircledTop{T,I}\end{tabular} &
  \begin{tabular}[c]{@{}c@{}}Wrong \\ treatment\end{tabular} \\ \hline
3 &
  GI Genius$^{\ddagger}$ &
  \begin{tabular}[c]{@{}c@{}}Gastroenterology/\\ Urology\end{tabular} &
  \begin{tabular}[c]{@{}c@{}}Gastro-\\ intestinal\\ lesion\\ detection\end{tabular} &
  \begin{tabular}[c]{@{}c@{}}Medical\\ practitioners\end{tabular} &
  \begin{tabular}[c]{@{}c@{}}Convolutional\\ neural\\ networks (CNN) $^*$\end{tabular} &
  Video &
  \begin{tabular}[c]{@{}c@{}}Amin et al.~\cite{amin2022two} \CircledTop{T}\end{tabular} &
  \begin{tabular}[c]{@{}c@{}}Wrong\\ diagnosis\end{tabular} \\ \hline
4 &
  SOZO$^{\ddagger}$ &
  \begin{tabular}[c]{@{}c@{}}Gastroenterology/\\ Urology\end{tabular} &
  \begin{tabular}[c]{@{}c@{}}Body fluid\\ analyzer for \\ assessing\\ protein-calorie\\ malnutrition\end{tabular} &
  \begin{tabular}[c]{@{}c@{}}Medical\\ practitioners\end{tabular} &
  CNN $^*$ &
  Numeric &
  Byra et al.~\cite{9251568} \CircledTop{I} &
  \begin{tabular}[c]{@{}c@{}}Wrong\\ diagnosis\end{tabular} \\ \hline
5 &
  \begin{tabular}[c]{@{}c@{}}WellDoc \\ BlueStar$^{\dagger}$\end{tabular} &
  General hospital &
  \begin{tabular}[c]{@{}c@{}}Diabetes \\ management\end{tabular} &
  \begin{tabular}[c]{@{}c@{}}Medical\\ practitioners,\\ patients\end{tabular} &
  \begin{tabular}[c]{@{}c@{}}Darknet-53\\ CNN\end{tabular} &
  Numeric &
  Lal et al.~\cite{lal2021adversarial} \CircledTop{I} &
  \begin{tabular}[c]{@{}c@{}}Wrong\\ diagnosis\end{tabular} \\ \hline
6 &
  d-Nav System$^{\dagger}$ &
  General hospital &
  \begin{tabular}[c]{@{}c@{}}Insulin dose\\ predictor\end{tabular} &
  \begin{tabular}[c]{@{}c@{}}Medical\\ practitioners,\\ patients\end{tabular} &
  \begin{tabular}[c]{@{}c@{}}Multi-layer\\ perception (MLP)\\ and LSTM\end{tabular} &
  Numeric &
  Zhou et al.~\cite{zhou2022robustness} \CircledTop{I} &
  \begin{tabular}[c]{@{}c@{}}Wrong\\ treatment\\ (Fatal)\end{tabular} \\ \hline
7 &
  \begin{tabular}[c]{@{}c@{}}MBT-CA \\ System$^{\ddagger}$\end{tabular} &
  Microbiology &
  Spectometry &
  \begin{tabular}[c]{@{}c@{}}Medical\\ practitioners\end{tabular} &
  DNN $^*$ &
  Numeric &
  \begin{tabular}[c]{@{}c@{}}Meiseles \\et al.~\cite{9313421} \CircledTop{I} \end{tabular}&
  \begin{tabular}[c]{@{}c@{}}Wrong\\ diagnosis\\ (Fatal)\end{tabular} \\ \hline
8 &
  KIDScore D3$^{\dagger}$ &
  \begin{tabular}[c]{@{}c@{}}Obstetrics \&\\ Gynaecology\end{tabular} &
  \begin{tabular}[c]{@{}c@{}}Embryo image\\ assessment\end{tabular} &
  \begin{tabular}[c]{@{}c@{}}Medical\\ practitioners\end{tabular} &
  \begin{tabular}[c]{@{}c@{}}Decentralized\\ federated learning\end{tabular} &
  Image &
  Nguyen et al.~\cite{nguyen2022novel} \CircledTop{P} &
  \begin{tabular}[c]{@{}c@{}}Wrong\\ diagnosis\end{tabular} \\ \hline
9 &
  \begin{tabular}[c]{@{}c@{}}NuVasive \\ Pulse System$^{\ddagger}$\end{tabular} &
  Orthopedic &
  \begin{tabular}[c]{@{}c@{}}Neurological\\ monitoring\end{tabular} &
  \begin{tabular}[c]{@{}c@{}}Medical\\ practitioners\end{tabular} &
  CNN $^*$ &
  Image &
  Joel et al.~\cite{joel2021adversarial} \CircledTop{I} &
  \begin{tabular}[c]{@{}c@{}}Mistake in\\ surgery (Fatal)\end{tabular} \\ \hline
10 &
  ABMD Software$^{\dagger}$ &
  Radiology &
  \begin{tabular}[c]{@{}c@{}}Bone \\ densitometer\end{tabular} &
  \begin{tabular}[c]{@{}c@{}}Medical\\ practitioners\end{tabular} &
  \begin{tabular}[c]{@{}c@{}}Inception-v3 \\ and \\ Densenet-121 $^*$\end{tabular} &
  Image &
  Bortsova et al.~\cite{bortsova2021adversarial} \CircledTop{I} &
  \begin{tabular}[c]{@{}c@{}}Wrong\\ diagnosis\end{tabular} \\ \hline
11 &
  \begin{tabular}[c]{@{}c@{}}Deep Learning\\ Image\\ Reconstruction$^{\dagger}$\end{tabular} &
  Radiology &
  \begin{tabular}[c]{@{}c@{}}X-ray\\ reconstruction\end{tabular} &
  \begin{tabular}[c]{@{}c@{}}Medical\\ practitioners\end{tabular} &
  ResNet-18 &
  Image &
  \begin{tabular}[c]{@{}c@{}}Menon et al.~\cite{menon2021covid} \CircledTop{T}\\ Paul et al.~\cite{paul2020mitigating} \CircledTop{I}\end{tabular} &
  \begin{tabular}[c]{@{}c@{}}Wrong\\ diagnosis\end{tabular} \\ \hline
12 &
  Air Next$^{\ddagger}$ &
  Anesthesiology &
  Spirometer &
  \begin{tabular}[c]{@{}c@{}}Medical\\ practitioners\end{tabular} &
  \begin{tabular}[c]{@{}c@{}}CatBoost\\ ResNet-50 $^*$\end{tabular} &
  Image &
  Vargas et al.~\cite{vargas2020understanding} \CircledTop{I} &
  \begin{tabular}[c]{@{}c@{}}Wrong\\ diagnosis\end{tabular} \\ \hline
13 &
  \begin{tabular}[c]{@{}c@{}}One Drop\\ Blood Glucose\\ Monitoring\\ System$^{\ddagger}$\end{tabular} &
  \begin{tabular}[c]{@{}c@{}}Clinical \\ Chemistry\end{tabular} &
  \begin{tabular}[c]{@{}c@{}}Diabetes \\ management\end{tabular} &
  Patients &
  \begin{tabular}[c]{@{}c@{}}MLP\end{tabular} &
  Numeric &
  \begin{tabular}[c]{@{}c@{}}Levy-Loboda\\ et al.~\cite{levy2022personalized} \CircledTop{I} \end{tabular} &
  \begin{tabular}[c]{@{}c@{}}Wrong\\ treatment\\ (Fatal)\end{tabular} \\ \hline
14 &
  \begin{tabular}[c]{@{}c@{}}OTIS 2.1 and\\ THiA Optical\\ Coherence\\ Tomography\\ System$^{\ddagger}$\end{tabular} &
  \begin{tabular}[c]{@{}c@{}}General and\\ Plastic Surgery\end{tabular} &
  \begin{tabular}[c]{@{}c@{}}Human tissue\\ imaging\end{tabular} &
  \begin{tabular}[c]{@{}c@{}}Medical\\ practitioners\end{tabular} &
  \begin{tabular}[c]{@{}c@{}}Support Vector\\ Machines\\ (SVM)\end{tabular} &
  Image &
  Ma et al.~\cite{ma2021understanding} \CircledTop{I} &
  \begin{tabular}[c]{@{}c@{}}Wrong\\ diagnosis\end{tabular} \\ \hline
15 &
  \begin{tabular}[c]{@{}c@{}}EarliPoint\\ System$^{\ddagger}$\end{tabular} &
  Neurology &
  \begin{tabular}[c]{@{}c@{}}Diagnosis of\\ Pediatric \\ Autism Spectrum\\ Disorder\end{tabular} &
  \begin{tabular}[c]{@{}c@{}}Medical\\ practitioners\end{tabular} &
  \begin{tabular}[c]{@{}c@{}}Graph Neural\\ Network\\ (GNN)\end{tabular} &
  Image &
  Chen et al.~\cite{chen2022adversarial} \CircledTop{T} &
  \begin{tabular}[c]{@{}c@{}}Wrong\\ diagnosis\end{tabular} \\ \hline
16 &
  BrainScope TBI$^{\ddagger}$ &
  Neurology &
  \begin{tabular}[c]{@{}c@{}}Brain injury\\ assessment\end{tabular} &
  \begin{tabular}[c]{@{}c@{}}Medical\\ practitioners\end{tabular} &
  \begin{tabular}[c]{@{}c@{}}CNN + \\ Recurrent neural \\ networks (RNN)\end{tabular} &
  Numeric &
  Yu et al.~\cite{yu2023perturbing} \CircledTop{I} &
  \begin{tabular}[c]{@{}c@{}}Wrong\\ treatment\\ (Fatal)\end{tabular} \\ \hline
17 &
  IDx-DR v2.3$^{\dagger}$ &
  Ophthalmic &
  \begin{tabular}[c]{@{}c@{}}Diabetic\\ Retinopathy\\ Detection\end{tabular} &
  \begin{tabular}[c]{@{}c@{}}Medical\\ practitioners\end{tabular} &
  \begin{tabular}[c]{@{}c@{}}Federated\\ learning\end{tabular} &
  Image &
  Nielsen et al.~\cite{nielsen2022investigating} \CircledTop{I} &
  \begin{tabular}[c]{@{}c@{}}Wrong \\ diagnosis\\ (loss of vision)\end{tabular} \\ \hline
18 &
  \begin{tabular}[c]{@{}c@{}}Iris Intelligent\\ Retinal Imaging\\ System$^{\dagger}$\end{tabular} &
  Ophthalmic &
  \begin{tabular}[c]{@{}c@{}}Storage,\\ management and\\ display of retinal\\ images\end{tabular} &
  \begin{tabular}[c]{@{}c@{}}Medical\\ practitioners\end{tabular} &
  DNN &
  Image &
  \begin{tabular}[c]{@{}c@{}}Mangaokar \\et al.~\cite{mangaokar2020jekyll} \CircledTop{I} \end{tabular}&
  \begin{tabular}[c]{@{}c@{}}Wrong \\ diagnosis\\ (loss of vision)\end{tabular} \\ \hline
19 &
  Paige Prostate$^{\dagger}$ &
  Pathology &
  \begin{tabular}[c]{@{}c@{}}Cancer\\ diagnosis\end{tabular} &
  \begin{tabular}[c]{@{}c@{}}Medical\\ practitioners\end{tabular} &
  CNN &
  Numeric &
  Hu et al.~\cite{hu2022adversarial} \CircledTop{T} &
  \begin{tabular}[c]{@{}c@{}}Wrong\\ treatment (Fatal)\end{tabular} \\ \hline
20 &
  \begin{tabular}[c]{@{}c@{}}Tissue of Origin\\ Test Kit$^{\ddagger}$\end{tabular} &
  Pathology &
  \begin{tabular}[c]{@{}c@{}}Malignant Tumor\\ diagnosis\end{tabular} &
  \begin{tabular}[c]{@{}c@{}}Medical\\ practitioners\end{tabular} &
  SVM &
  Image &
  Ma et al.~\cite{ma2021understanding} \CircledTop{I} &
  \begin{tabular}[c]{@{}c@{}}Wrong\\ treatment (Fatal)\end{tabular} \\ \hline
\end{tabular}
\setlength\tabcolsep{6pt}
\caption{A study of different FDA-Approved ML-enabled medical devices and their security vulnerabilities\\{\footnotesize
$\dagger$: Software as medical device, $\ddagger$: Software in medical device, $^*$: Best-guessed ML algorithm,\\ \CircledTop{T}: Training-time attack, \CircledTop{I}: Inference-time attack, \CircledTop{P}: Privacy attack}}
\label{tab:knownmlattacks}
\end{table*}


We investigate how an adversary can exploit the vulnerabilities identified in \S\ref{sec:knownattacks}. Table~\ref{tab:knownmlattacks} shows that all the identified ML attacks involve manipulating inputs during training or inference. For each ML-enabled device, we search for compatible peripheral devices and communication channels that would allow adversaries to introduce malicious data into the ML engine. We also assess the adequacy (or the lack thereof) of manufacturers' risk assessments, as mentioned in their Premarket Notifications, to determine their effectiveness in preventing such security risks. Table~\ref{tab:vulnerable-devices} presents this study.


\gmit{\noindent\textbf{Survey Process. } To understand the vulnerability landscape of each device, we performed the following two steps.

\noindent\textit{Step 1. Identifying peripheral devices and communication media compatible with the ML-enabled device:} We identify compatible peripheral sensor devices, communication media, and operating system from its Premarket Notification summary~\cite{fdaml} and information on the manufacturer's website. One or more of these can be a potential point of attack.

\noindent\textit{Step 2. Identifying vulnerable peripherals that can be exploited for attacking the ML engine: }For each potential attack point, we look for known attacks and vulnerabilities by searching research papers and vulnerability databases~\cite{mitre-cve,nist-cve}. 
We list at least one vulnerability that would allow an adversary to eavesdrop or inject malicious data into the ML engine, enabling them to execute the attacks identified in \S\ref{sec:knownattacks}.
This list of vulnerabilities is not comprehensive. We highlight at least one vulnerability to motivate the risk analysis technique.
}

\noindent\textbf{Insights. }We summarize the insights from this study below. 
\begin{enumerate}[leftmargin=*]
    \item We found known vulnerabilities in the peripheral devices compatible with several ML-enabled devices. While most vulnerabilities affect only a small group of devices
    , a few vulnerabilities affect all devices of a certain type. For instance, the Conexus telemetry protocol vulnerability~\cite{ecgvul1} only affects the ECG monitors from Medtronic. However,  another attack~\cite{wang2021can} affects all infrared-sensitive cameras.
    \item Some vulnerabilities (e.g., \cite{ecgvul1} for the Cardiologs ECG Analysis Platform, and \cite{wang2021can} for Oxehealth Vital Signs) require the adversary to execute the attack locally as they have to be within the Bluetooth~\cite{ecgvul1} communication range, or within the range of infrared light emission~\cite{wang2021can}. Such attacks can be executed by insiders or by breaching the physical security of the hospital or the patient's home.
    \item Many of the vulnerabilities can be exploited remotely (e.g., \cite{ir1} for Oxehealth and \cite{window2} for the IDx-DR) 
    over the Internet. 
    Since connectivity to the Internet is mandatory for these devices, preventing remote attacks is challenging.
    \item In some cases, identifying the attack path is challenging. For example, the IDx-DR software relies on inputs from the Topcon NW200 Fundus camera. Although we found no known vulnerability in the camera, it comes bundled with a computer running Windows 7 by default, which has known vulnerabilities~\cite{window2}. These Windows 7 vulnerabilities could enable an adversary to inject malicious inputs into the ML engine. While updates for Windows 7 may address such vulnerabilities, medical devices typically do not receive routine security updates. In a specific case~\cite{window2}, the vendor even decided not to release a patch, assuming most users would upgrade to Windows 10.
    \item We did not find any known vulnerability in the peripheral devices for some systems (e.g., SOZO, WellDoc Bluestar, and Air Next). However, many of these systems use Bluetooth, Internet communications, and web services. Adversaries can exploit vulnerabilities~\cite{rasmussen2022blurtooth
    } in these communication channels for injecting adversarial inputs.
    \item We found that many of the ML-enabled device manufacturers (e.g., the NuVasive Pulse System) do not perform any security evaluation, and only focus on accuracy and safe operating conditions (e.g., protecting the devices from electrical hazards). 
    Even the manufacturers who consider security, rarely consider the peripheral devices.
    For instance, the developers of the IDx-DR software evaluate the software for various security risks, but not its peripheral device, the Topcon NW200 Fundus Camera. However, security evaluation of the software alone is insufficient. This is because an adversary might execute the inference-time attack~\cite{nielsen2022investigating} shown in Table~\ref{tab:knownmlattacks} by exploiting the vulnerability in the camera~\cite{window2} to install malware that manipulates the images that are sent to the input of the ML engine.  
    \end{enumerate}



\begin{table*}[hbt!]
\footnotesize
\setlength\tabcolsep{1.5pt}
\begin{tabular}{|c|c|c|l|c|}
\hline
\rowcolor[HTML]{EFEFEF} 
\textbf{\begin{tabular}[c]{@{}c@{}}Sl.\\ No.\end{tabular}} &
  \textbf{\begin{tabular}[c]{@{}c@{}}Device \\ Name\end{tabular}} &
  \textbf{\begin{tabular}[c]{@{}c@{}}Risk assessment guideline followed\end{tabular}} &
  \multicolumn{1}{c|}{\cellcolor[HTML]{EFEFEF}\textbf{\begin{tabular}[c]{@{}c@{}}Known attacks and\\vulnerabilities in compatible\\ peripheral sensor devices\end{tabular}}} &
  \textbf{\begin{tabular}[c]{@{}c@{}}Connected to\\ the Internet\\ or Bluetooth ?\end{tabular}} \\ \hline
1 &
  \begin{tabular}[c]{@{}c@{}}CardioLogs ECG Analysis\\ Platform$^{\dagger}$\end{tabular} &
  \multicolumn{1}{l|}{\begin{tabular}[c]{@{}l@{}}Inadequate information - Acknowledges the need for \\ cybersecurity of cloud-based software\end{tabular}} &
  \begin{tabular}[l]{@{}l@{}}Portable ECG Monitors - \{\cite{ecgvul1
  }\} \CircledTop{L}\end{tabular} & \begin{tabular}[c]{@{}c@{}}Cellular network,\\Bluetooth \end{tabular}
   \\ \hline
2 &
  Oxehealth Vital Signs$^{\dagger}$ &
  \multicolumn{1}{l|}{\begin{tabular}[c]{@{}l@{}}Guidance for the Content of Premarket Submissions for\\Management of Cybersecurity in Medical Devices~\cite{fdapremarket}\end{tabular}} &
  \begin{tabular}[l]{@{}l@{}}Infra-red sensitive cameras\\- \cite{wang2021can} \CircledTop{L}, \{\cite{ir1
  }\} \CircledTop{R} \end{tabular}& Intranet / Internet
   \\ \hline
3 &
  GI Genius$^{\ddagger}$ &
  \multicolumn{1}{l|}{\begin{tabular}[c]{@{}l@{}}Moderate  level of concern as defined in the\\``Guidance for the Content of Premarket Submissions\\for Software Contained in Medical Devices.''~\cite{fdapremarket}\end{tabular}} &
  \begin{tabular}[l]{@{}l@{}}
  Endoscope cameras - \{\cite{endo-1
  }\} \CircledTop{R} 
  \end{tabular}& Intranet / Internet
   \\ \hline
4 &
  SOZO$^{\ddagger}$ &
  None &
  \begin{tabular}[l]{@{}l@{}}No third-party peripheral device used\end{tabular} & \begin{tabular}[l]{@{}l@{}}Bluetooth,\\Intranet/Internet\end{tabular}
   \\ \hline
5 &
  WellDoc BlueStar$^{\dagger}$ &
  \begin{tabular}[l]{@{}l@{}}Guidance for the Content of Premarket Submissions for\\Management of Cybersecurity in Medical Devices~\cite{fdapremarket}\end{tabular} &
  \begin{tabular}[l]{@{}l@{}}No vulnerability identified in peripheral\\sensor devices\end{tabular} & \begin{tabular}[l]{@{}l@{}}
  Bluetooth,\\
  Cloud Service API
  \end{tabular}
   \\ \hline
6 &
  d-Nav System$^{\dagger}$ &
  \begin{tabular}[l]{@{}l@{}}Guidance for the Content of Premarket Submissions for\\Management of Cybersecurity in Medical Devices~\cite{fdapremarket}\end{tabular}&
  \begin{tabular}[l]{@{}l@{}}No vulnerability identified in peripheral\\sensor devices\end{tabular} & Cloud Service API
   \\ \hline
7 &
  MBT-CA System$^{\ddagger}$ &
  None & \begin{tabular}[l]{@{}l@{}}No third-party peripheral device used\end{tabular} &
  No \\ \hline
8 &
  KIDScore D3$^{\dagger}$ &
  None & \begin{tabular}[l]{@{}l@{}}No vulnerability identified in peripheral\\sensor devices\end{tabular}
   & Intranet / Internet
   \\ \hline
9 &
  NuVasive Pulse System$^{\ddagger}$ &
  None & \begin{tabular}[l]{@{}l@{}}Infra-red sensitive cameras\\- \cite{wang2021can} \CircledTop{L}, \{\cite{ir1
  }\} \CircledTop{R} \end{tabular}& Internet\\\hline
10 &
  ABMD Software$^{\dagger}$ &
  None & \begin{tabular}[l]{@{}l@{}}No vulnerability identified in peripheral\\sensor devices\end{tabular}
   & Unknown
   \\ \hline
11 &
  \begin{tabular}[c]{@{}c@{}}Deep Learning\\ Image Reconstruction$^{\dagger}$\end{tabular} &
  None & \begin{tabular}[l]{@{}l@{}}
  X-ray machines 
  - \{\cite{xray-1
  }\} \CircledTop{R} \end{tabular}& Unknown
   \\ \hline
12 &
  Air Next$^{\ddagger}$ &
  None &
  \begin{tabular}[l]{@{}l@{}}No third-party peripheral device used\end{tabular} &
  \begin{tabular}[c]{@{}c@{}}Bluetooth, \\ Internet\end{tabular} \\ \hline
13 &
  \begin{tabular}[c]{@{}c@{}}One Drop Blood Glucose\\ Monitoring System\end{tabular} &
  None &
  \begin{tabular}[l]{@{}l@{}}No third-party peripheral device used\end{tabular} &
  \begin{tabular}[c]{@{}c@{}}Bluetooth, \\ Internet\end{tabular} \\ \hline
14 &
  \begin{tabular}[c]{@{}c@{}}OTIS 2.1 and THiA Optical\\ Coherence Tomography System\end{tabular} &
  \begin{tabular}[l]{@{}l@{}}
  ANSI AAMI ISO 14971:2007/(R)2010~\cite{aami2007ansi},\\
  IEC 62304:2006/A1:2015~\cite{international2015iec}
  \end{tabular}& \begin{tabular}[l]{@{}l@{}}No third-party peripheral device used\end{tabular}
   & Unknown
   \\ \hline
15 &
  EarliPoint System$^{\ddagger}$ &
  None &
  \begin{tabular}[c]{@{}l@{}}Webcams installed on personal\\computers - \cite{webcam} \CircledTop{R}\end{tabular} & Internet
   \\ \hline
16 &
  BrainScope TBI$^{\ddagger}$ &
  None & No third-party peripheral devices used
   & Internet
   \\ \hline
17 &
  IDx-DR v2.3$^{\dagger}$ &
  \multicolumn{1}{l|}{\begin{tabular}[c]{@{}l@{}}Considers security concerns related to data \\ confidentiality, integrity, availability, denial of service\\attacks and malware. Risks related to the failure of various\\software components and their potential impact on\\patient reports were also adequately addressed~\cite{idx}.\end{tabular}} &
  \begin{tabular}[l]{@{}l@{}}This device uses the Topcon NW200\\Fundus camera, which comes packaged\\with a PC running Windows 7 OS. The \\Windows 7 OS has known\\vulnerabilities.
  - \{\cite{window2
  }\} \CircledTop{R}
  \end{tabular} & Internet
   \\ \hline
18 &
  \begin{tabular}[c]{@{}c@{}}Iris Intelligent Retinal Imaging\\ System$^{\dagger}$\end{tabular} &
  \multicolumn{1}{l|}{Ensures HIPAA~\cite{annas2003hipaa} compliance} &
  \begin{tabular}[l]{@{}l@{}}Retinal cameras such as Topcon\\NW200 
  - Same vulnerable peripherals\\as in the case of IDx-DR v2.3
  \end{tabular}& Internet
   \\ \hline
19 &
  Paige Prostate$^{\dagger}$ &
  \multicolumn{1}{l|}{\begin{tabular}[c]{@{}l@{}}Considers software security 
  as per ``Content of Premarket \\ Submissions for Management of Cybersecurity in \\ Medical Devices: Guidance for Industry and Food and\\Drug Administration Staff". Also encrypts the \\communication between the device and servers.\end{tabular}} &
   \begin{tabular}[l]{@{}l@{}}
   Medical scanners 
   - \{\cite{phillips1
   }\} \CircledTop{L}
   \end{tabular}& Internet
   \\ \hline
20 &
  Tissue of Origin Test Kit$^{\ddagger}$ &
  None & \begin{tabular}[l]{@{}l@{}}No third-party peripheral device used\end{tabular}
   & Internet
   \\ \hline
\end{tabular}
\caption{Known vulnerabilities in peripheral devices and communication media compatible with FDA-approved ML-enabled medical devices. \CircledTop{L}: Locally exploitable only, \CircledTop{R}: Remotely exploitable}
\label{tab:vulnerable-devices}
\setlength\tabcolsep{6pt}
\end{table*}

\section{Case Study}\label{sec:case-study}
We present a case study to 
demonstrate the security risks in the ML-enabled BGMS described in Section \ref{sec:motiv-bgms}. We show a practical attack on the BGMS in which the attacker exploits the vulnerabilities in connected devices to negatively affect the predictions of the ML-enabled decision-making component \footnote{\eln{No human subjects were used in our experiments. Instead, we rely on a publicly available anonymized dataset and a publicly available prediction model that resembles the original model in terms of functionality and features.}}. 

\subsection{Attack Description}

\noindent\textbf{Adversarial Goal.} 
The attacker aims to endanger a targeted patient's life by causing the ML model to misdiagnose the patient's condition, thereby leading to an incorrect insulin dose suggestion. While minor prediction errors are benign, a substantial error could have life-threatening consequences. In this case study, we consider an attacker aiming to make the model predict a high blood glucose level (hyperglycemia) when the patient actually has a low (hypoglycemia) or normal blood glucose level. If the attacker succeeds, the BGMS would erroneously recommend more insulin, causing the patient's glucose level to drop significantly below normal. The implications range from incorrect diagnosis (e.g., in the WellDoc BlueStar system) to incorrect treatment (e.g., in d-Nav and One Drop BGMS systems), potentially leading to fatal outcomes.


\noindent\textbf{Adversarial Capabilities.} 
We assume the attacker has reasonable and realistic capabilities,
wherein they can only tamper with the CGM measurements. Manipulating the manually entered finger-based glucose readings, carbohydrate intake, and bolus dose is beyond the attacker's capabilities. \eln{However, the attacker can compromise the smartphone \cite{suarez2015compartmentation} to gain read-only access to these values once they have been gathered from external sensors or after being manually entered by the patients in the mobile app.} The attacker is oblivious to the structure and parameters of the underlying ML model (black box attack), and does not have access to the training set.
The attacker can attack the Bluetooth communication stack via known exploits \cite{rasmussen2022blurtooth} to intercept and manipulate the CGM measurements. This is because the FDA-approved diabetes management devices (e.g., One Drop and WellDoc BlueStar) use Bluetooth communication to transmit the collected measurements. 


\noindent\textbf{Attack Strategy.} The attacker aims to misdiagnose the patient as hyperglycemic by pushing predicted blood glucose levels toward the hyperglycemic range. This involves modifying hypoglycemic or normal blood glucose levels to values exceeding 125 mg/dL (hyperglycemic while fasting) or 180 mg/dL (hyperglycemic postprandial). To achieve this, the attacker manipulates CGM readings for a specific duration, causing the BGMS app to misdiagnose the patient's blood glucose level. Determining the minimum time duration and extent of manipulation requires careful consideration. 


\subsection{Experimental Setup}
We first present the ML model used in the BGMS setup, followed by a description of the dataset used for our experiments.
Next, we describe the Universal Robustness Evaluation Toolkit (URET)~\cite{eykholturet}, used for generating the adversarial inputs.

\noindent\textbf{Targeted ML model.} Since the specific ML algorithm used in the Dreamed Advisor Pro app (described in Section \ref{sec:motiv-bgms}) is confidential, we approximated it using a time-series prediction model developed by Rubin-Falcone et al. \cite{rubin2020deep}. This model uses a bidirectional long short-term memory (LSTM) recurrent neural network (RNN) architecture, 
and uses root mean square error (RMSE) and mean absolute error (MAE) to evaluate the prediction accuracy. Intuitively, both RMSE and MAE indicate the difference between predicted and actual glucose levels. The higher the difference, the worse the prediction. Further, our chosen target model uses a neural network similar to the FDA-approved d-Nav System \cite{dnav} in Tables \ref{tab:knownmlattacks} and \ref{tab:vulnerable-devices} (i.e., LSTM).

Rubine-Falcone et al. \cite{rubin2020deep} 
built two models - (i) a personalized model for each patient trained on the patient's data, and (ii) an aggregate model trained on the data of all patients. Their average RMSEs are 18.2 and 31.7, and average MAEs are 12.8 and 23.6 on the 30 and 60-minute horizons, respectively.


\noindent\textbf{Dataset.} To demonstrate the impact of adversarial inputs on the predictions of the targeted ML model, we use the 2020 OhioT1DM dataset~\cite{marling2020ohiot1dm}. 
This was used by the target model developers~\cite{rubin2020deep} for evaluating its accuracy. 
The dataset comprises physiological measurements of six Type-1 diabetic patients. 
The main features are CGM blood glucose measurements, finger-based measurements, basal insulin, bolus dose, carbohydrate intake, heart rate, sleeping patterns and acceleration, besides other physiological, and self-reported life-event features. The dataset spans eight weeks and consists of $\approx$10000 samples for training, and 2500 for testing, both recorded at approximately 5-minute intervals per patient.



\noindent\textbf{Universal Robustness Evaluation Toolkit (URET).}
In Table \ref{tab:knownmlattacks}, we found three diabetes management devices (5,6, and 13) that are vulnerable to inference-time attacks. Hence, we decided to launch an \emph{evasion attack} against our target model, which manipulates data points at inference time by introducing slight perturbations to their original values to evade detection by the ML model and cause misclassification \cite{biggio2013evasion}. We use URET~ \cite{eykholturet}, a general-purpose framework for generating adversarial inputs for evasion attacks. Unlike most evasion attack techniques 
that focus on the image classification domain \cite{finlayson2019adversarial, finlayson2018adversarial, carlini2017towards, paul2020mitigating}, URET can generate adversarial inputs regardless of the data types or task domain. 

URET takes in a benign input instance 
and a set of pre-defined input transformations and attempts to find a sequence of transformations
(e.g., increment or replace glucose level) resulting in an adversarial input that is both semantically and functionally correct.  The URET framework is compatible with a wide variety of data types and domains and allows the user to specify which feature values to manipulate, and bounds and constraints on the feature values. As URET allows specifying constraints on the feature values, 
we can ensure the generated adversarial samples abide by victims' physiological limits. 

Since we assume the attacker can only modify CGM values, we specify the CGM feature indices in URET's configuration file as the only modifiable feature. 
To ensure that adversarial CGM values respect the physiological levels, we constrain them to be between 125 and 499 mg/dL for fasting levels, since a hyperglycemic glucose level in a fasting state should exceed 125 mg/dL, and between 180 and 499 mg/dL for postprandial levels, since a hyperglycemic glucose level in a postprandial state should exceed 180 mg/dL (499 mg/dL is the highest reported glucose level in the OhioT1DM dataset). Further, since there are missing physiological measurements at specific timestamps in the dataset, especially CGM measurements, we specify the relationship between CGM glucose and a feature called ``missing'' as a dependency. 
This ensures the ML model does not 
interpret the missing CGM values as zero values. 

In addition, the attacker can observe other feature values (e.g., bolus dose), which in turn play a role in adversarial data generation, since the attacker needs read access to those features to use the loss function to rank the adversarial input.

\subsection{Results}
\begin{figure}[t]
\centering
\includegraphics[width=0.35\textwidth]{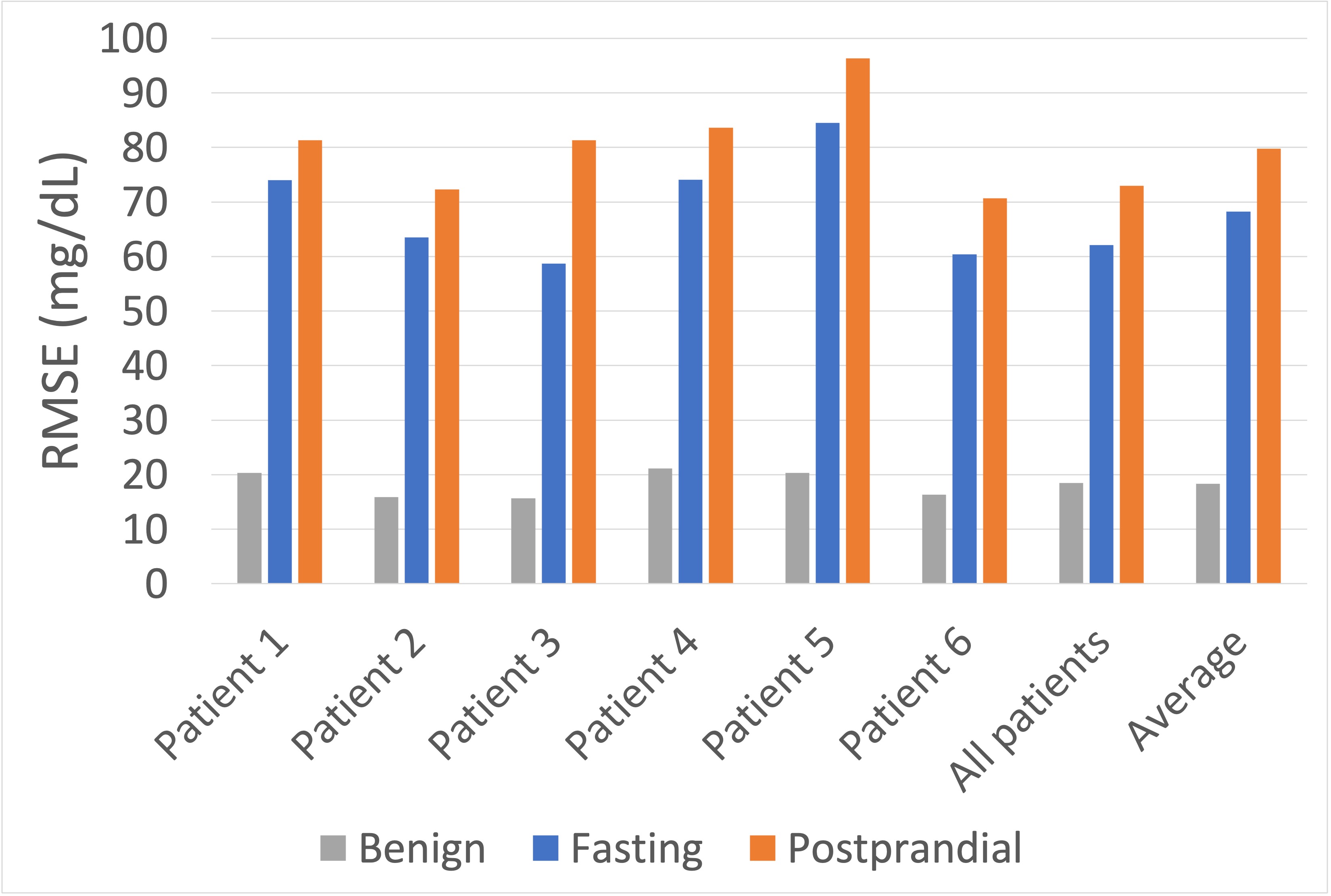}
\caption{RMSE of the benign model, the model attacked with fasting hyperglycemic blood glucose levels, and with postprandial hyperglycemic blood glucose levels.}
\label{fig:RMSE Results}
\end{figure}

While the target models use RMSE and MAE to evaluate the prediction accuracy, we consider the RMSE results only, as the two metrics show similar behaviors. Figure \ref{fig:RMSE Results} shows the performance of personalized patient models and the aggregate model trained on all patients' data (`All patients') before and after the URET attacks while fasting and postprandial. The maximum RMSE value for the benign models is around 21 mg/dL, while the average RMSE across all 7 benign models is around 18.3 mg/dL. The figure also shows the RMSE values after the attack for both fasting ($>$125 mg/dL) and postprandial ($>$180 mg/dL) hyperglycemic glucose levels. Thus, 
the RMSE values increase regardless of the used threshold since glucose values are driven further away from the actual values. 

Moreover, postprandial RMSE values are consistently higher than fasting RMSE values due to the use of a higher threshold for the lower bound of the adversarial input (i.e., 180 mg/dL instead of 125 mg/dL). The hike in RMSE values between the benign and the attacked models in both cases shows a significant difference between the actual and predicted blood glucose levels, implying diminished accuracy of the ML model. 
This would cause a potentially fatal insulin overdose. 

\begin{figure}[t]
\centering
\includegraphics[width=0.35\textwidth]{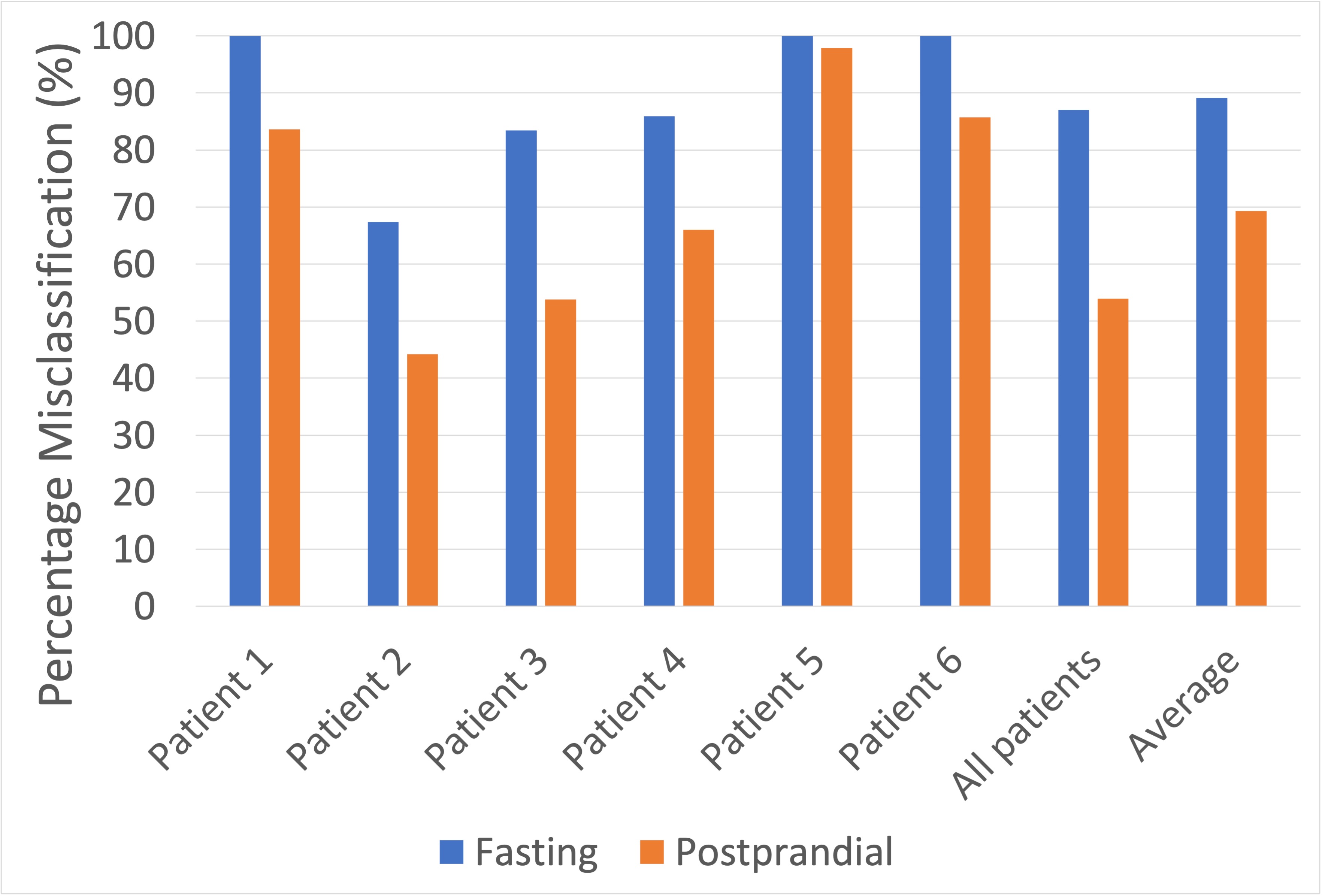}
\caption{Percentage of originally normal glucose instances that are misclassified as hyperglycemic.}
\label{fig:Normal to Hyper}
\end{figure}

\begin{figure}[t!]
\centering
\includegraphics[width=0.35\textwidth]{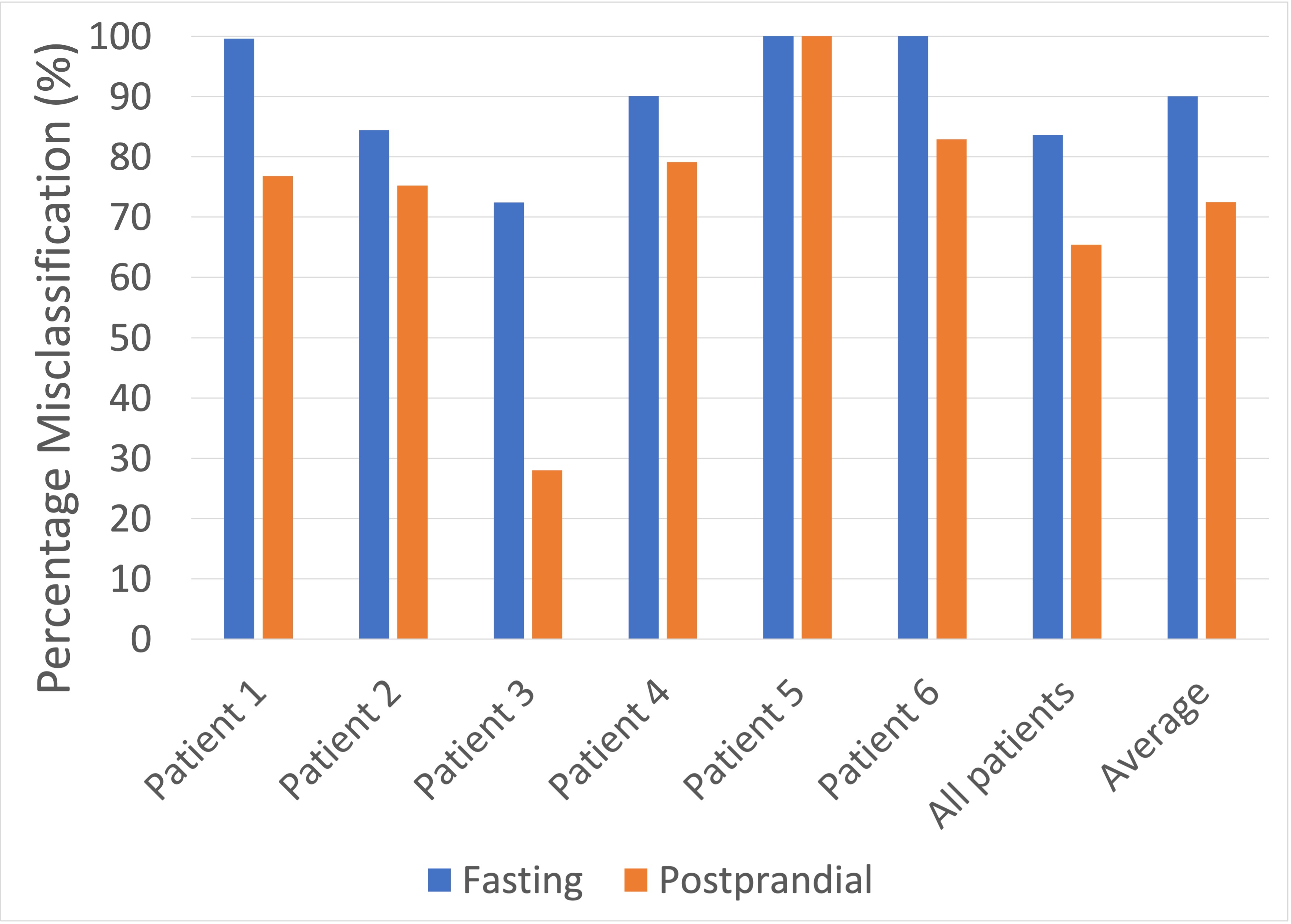}
\caption{Percentage of originally hypoglycemic glucose instances that are misclassified as hyperglycemic.}
\label{fig:Hypo to Hyper}
\end{figure}

Figures \ref{fig:Normal to Hyper} and \ref{fig:Hypo to Hyper} show the percentage of instances that are misdiagnosed as hyperglycemic while actually being normal and hypoglycemic, respectively, for both the fasting and postprandial attack scenarios. A higher misclassification percentage demonstrates more susceptibility to the respective attack scenario, while a lower percentage implies more difficulty in crafting successful attacks.
We make three observations from the two figures. 
\textit{First,} URET achieves considerably high attack success rates, reaching up to 100\% in some cases. 
URET achieves comparable attack success rates in both normal-to-hyperglycemic and hypoglycemic-to-hyperglycemic scenarios on average, demonstrating the robustness of the attack generated by URET across different initial glucose levels.
\textit{Second, }the attack success rate is consistently higher during fasting compared to postprandial states, indicating that attacking a fasting patient is relatively easier for the adversary. This is because a smaller amount of perturbation is performed when increasing the CGM glucose from the original value to fasting hyperglycemic levels as opposed to postprandial hyperglycemic levels. This confirms that URET is more successful when the perturbation margin is smaller. \textit{Third,} patients exhibit varying resilience to the URET attacks. For example, Figure \ref{fig:Normal to Hyper} shows that the success rate of misdiagnosing a patient from normal to hyperglycemic is the lowest for patient 2 (67.4\% fasting, and 44.2\% postprandial) and is the highest for patient 5 (100.0\% fasting, and 97.9\% postprandial) suggesting that it is more challenging for URET to attack patient 2 compared to patient 5. Similarly, Figure \ref{fig:Hypo to Hyper} shows that the attack success rate of misdiagnosing a patient from hypoglycemic to hyperglycemic is the lowest with patient 3 (72.4\% fasting, and 28.0\% postprandial) and is the highest with patient 5 (100.0\% fasting, and 100.0\% postprandial) suggesting that patient 5 is more vulnerable compared to patient 3. We hypothesize that this is because the rich history of some patients leads to a more robust prediction model that is more difficult to attack.

\textbf{Summary.} We demonstrated an attack where an attacker compromises the Bluetooth communication between the CGM peripheral device and the BGMS mobile app, enabling them to manipulate the measured glucose levels. The experimental results show that 
the attacker can cause the ML model to misdiagnose the patient's medical condition as hyperglycemic. In the best-case, this leads to a wrong diagnosis, and in the worst-case , it has fatal consequences for the patient.

\eln{
\subsection{Discussion: Attack Practicality and Broader Impact}
The assumptions made in the proposed attack are practical and align with previously demonstrated attacks \cite{rasmussen2022blurtooth}. While no real attack on a BGMS has been reported, security breaches in other medical devices, such as pacemakers, have been reported in the recent past \cite{Hern_2017}. The severity of these breaches is evident from the widespread recall of multiple pacemaker models and the substantial financial loss incurred by the manufacturer \cite{Lexology_2023}. These incidents also raised concerns about targeted attacks on high-profile individuals using such devices. For example, in 2013, former US Vice President Dick Cheney revealed that he had the wireless capabilities of his implanted pacemaker deactivated due to fears that an adversary could cause a cardiac arrest by sending a malicious signal to his pacemaker \cite{Gupta_2013}.

}

\section{Risk Assessment Techniques Evaluation}\label{sec:existingframeworks}
We assess the state-of-the-art risk assessment techniques, investigating both their strengths and limitations. We discuss the benefits of these methods in risk analysis processes while pointing out their shortcomings using the BGMS attack example. Finally, we highlight the need for a new risk assessment framework for ML-enabled connected medical systems.

\subsection{A Survey of Existing Techniques}\label{sec:survey-techniques}

We consider five broad categories of risk assessment methods. However, in the context of ML-enabled connected medical devices, these techniques often lack in three key aspects: (1) impact on affected users (more users affected, higher risk), (2) ease of vulnerability detection (easier detection, lower risk), and (3) ease of post-exploitation mitigation, including available remediation levels and responsible entities. 
Table \ref{tab:table_1} presents the extent to which existing risk assessment methods incorporate these factors. We describe the methods below. 

\noindent \textbf{DREAD}. 
DREAD~\cite{meier2003improving} is a \textit{risk rating system} built by Microsoft, primarily for evaluating risks posed to conventional web applications based on their damage potential, reproducibility, exploitability, affected users, and discoverability. However, the DREAD system cannot be used for ML-enabled connected medical systems. This is because DREAD does not detail how security risks in individual connected components may pose a threat to the overall system.


\noindent\textbf{STRIDE.} 
STRIDE~\cite{meier2003improving} by Microsoft is another qualitative model designed to \textit{identify and categorize threats} in web applications. It addresses six attack types: Spoofing, Tampering, Repudiation, Information Disclosure, Denial of Service, and Elevation of Privilege, and offers corresponding countermeasures. However, using STRIDE for an end-to-end risk assessment in ML-based medical systems is challenging because it does not account for adversaries exploiting peripheral devices. 



\noindent\textbf{FMEA.} Failure Modes and Effects Analysis (FMEA)~\cite{liu2020failure} is a foundational analytical technique to \textit{detect and mitigate} potential risks. This method involves a detailed examination of system components to identify potential causes of failure and their impact on system stability. However, it fails to provide end-to-end risk assessment for connected ML-based medical systems because it primarily focuses on individual component failures, overlooking broader systemic implications and how vulnerabilities propagate throughout the entire pipeline.

\renewcommand{\tabcolsep}{1.5pt}
\renewcommand{\arraystretch}{0.5}
\begin{table}[t]
\footnotesize
    \centering
    \begin{tabular}{lccccc}\toprule
Method         & \multicolumn{1}{c}{\begin{tabular}[c]{@{}c@{}}Damage \\ Potential\end{tabular}}  &  Exploitability  &  \multicolumn{1}{c}{\begin{tabular}[c]{@{}c@{}}Affected \\ Users\end{tabular}}  & \ Detectability  &  \multicolumn{1}{c}{\begin{tabular}[c]{@{}c@{}}Ease of \\ Mitigation\end{tabular}} \\ \midrule
DREAD   & \ding{51} & \ding{51} & \ding{51} & \ding{51} & \ding{55}\\\\
STRIDE   & \ding{51} & \ding{55} & \ding{55} & \ding{55} & \ding{55}\\\\
FMEA   & \ding{51} & \ding{55} & \ding{55} & \ding{51} & \ding{55}\\\\
CVSS   & \ding{51} & \ding{51} & \ding{55} & \ding{55} & \ding{55}\\\\
\begin{tabular}[l]{@{}l@{}}Other Works\\ \cite{mahler2020new,yaqoob2019integrated}\end{tabular}& \ding{51} & \ding{51} & \ding{55} & \ding{55} & \ding{55}\\\\
\end{tabular}
    \caption{Comparison of factors considered in existing risk analysis frameworks}
    \label{tab:table_1}
\end{table}
\renewcommand{\tabcolsep}{6pt}
\renewcommand{\arraystretch}{1}

\noindent\textbf{CVSS.} The Common Vulnerability Scoring System (CVSS)~\cite{abraham2015common} is an open risk scoring framework designed by FIRST.Org, Inc. to \textit{capture the severity level} of software vulnerabilities. CVSS comprises three metric groups: Base, Temporal, and Environmental, capturing different vulnerability characteristics. While it is not a risk assessment model \cite{abraham2015common}, it helps prioritize threats across system components through context-specific choices made by the risk management team.
However, it lacks consistency in prioritizing metrics, notably in ML-enabled medical devices where availability loss could be life-threatening, unlike in web applications where it might lead to user dissatisfaction or financial repercussions \cite{spring2021time}.

\noindent\textbf{FDA-approved Security Standards for Medical Devices.} 
The FDA has established a cybersecurity guideline to help the industry identify cybersecurity risks in medical devices \cite{fdapremarket}. However, their primary focus is on risks associated with communication between medical devices and IT networks. Consequently, addressing vulnerabilities in AI/ML-enabled medical devices that do not interface with an IT network remains a challenge.

\textbf{Academic Research Efforts and Industry.} There has been a rich literature on performing risk assessment within the healthcare sector \cite{mahler2020new,yaqoob2019integrated}. 
While companies typically conduct preliminary risk evaluations for their ML models, they do not analyze the security risks associated with deploying the models in a connected healthcare environment \cite{wu2021medical}. Rather, their concern is ensuring the ML model is trained on an unbiased dataset, and evaluated using a diverse dataset.  \halla{Similarly, while there are companies that claim to offer penetration testing services for medical devices, their testing is conducted in-silo rather than on the end-to-end connected system.}


\subsection{Limitations of Existing Techniques} 
We use the ML-enabled BGMS as an example to highlight the shortcomings of existing techniques in adequately considering all the factors required for an ideal risk analysis. 


\par \noindent\textbf{DREAD.} DREAD fails to assess the actual potential damage if a peripheral device like the glucose meter or smartphone is compromised. In such cases, the attacker can not only access the data but also to manipulate it, thereby compromising both the integrity and confidentiality of the system. 
However, DREAD cannot differentiate between them accurately.



\par \noindent \textbf{STRIDE.} 
An attacker can manipulate data in several ways before it is fed into the ML engine. Nevertheless, according to the STRIDE model, all these methods are categorized under Tampering, without any additional insights into how these manipulations were performed.

\par \noindent\textbf{FMEA}. FMEA could potentially assess the risks associated with individual peripheral devices in the BGMS, like glucose meters or smartphones. However, FMEA does not offer insights into the risks associated with the propagation of this vulnerability within the system or its potential impact on the ML model's predictions.

\noindent \textbf{CVSS.} When a vulnerability impacts the availability of the BGMS, the CVSS assigns the same risk level as it would for a similar incident in other domains like web applications. This approach is not appropriate, as an availability issue in the BGMS could potentially lead to irreversible harm to patients.
\par \noindent\textbf{FDA-approved Security Standards for Medical Devices.} Encountering physical attacks like electromagnetic interference \cite{mortazavi2014electromagnetic} is independent of the connectivity between the glucose meter and the IT network. Hence, the FDA-approved security standards also fall short of an ideal risk analysis.

\subsection{Need for a New Risk Assessment Framework}
\begin{figure}[t]
    \centering
    \includegraphics[scale=0.1]{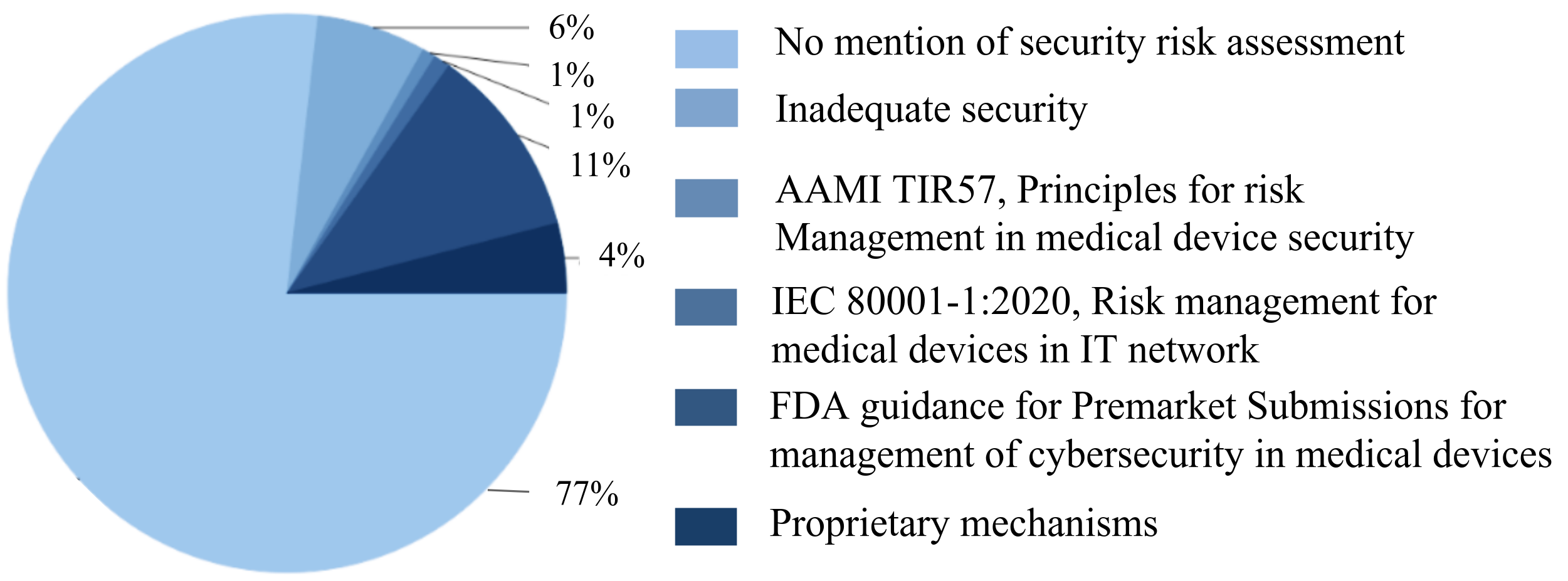}
    \caption[justification=centering]{Security risk assessment techniques by manufacturers of FDA-approved ML-enabled medical systems based on~\cite{fdapremarket}}
    \label{fig:riskassessment}
\end{figure}


Our investigation into the types of security risk assessment performed by manufacturers of ML-enabled medical devices is summarized in Figure~\ref{fig:riskassessment}.We find that over 80\% of these manufacturers either do not provide information about the assessment in their documentation, or employ inadequate assessment methods. Another 5\% use proprietary mechanisms, making it challenging to assess the adequacy of their approach. The remaining 12\% utilize existing risk assessment techniques, which, as discussed, are insufficient for risk assessment of ML-enabled connected medical systems. Therefore, developing an efficient risk assessment technique for ML-enabled medical devices remains an open challenge. 




\section{Conclusion and Future Research Directions}
We conducted a detailed study of security risks associated with modern AI/ML-enabled medical devices, stemming from vulnerabilities in connected peripheral devices. We conducted a systematic security analysis of FDA-approved commercial AI/ML-enabled devices. Our analyses reveal vulnerabilities of these devices to existing adversarial attacks, raising concerns about the suitability of using such safety-critical devices on real-world patients. To validate our analysis, we executed a realistic adversarial attack on an ML-enabled blood glucose monitoring system, identifying security risks in the process. Additionally, we studied state-of-the-art risk assessment frameworks, underscoring their limitations in identifying security risks in connected ML-enabled medical systems and highlighting the need for a new framework. 

Our work opens up three interesting future work directions -- \textbf{(1)} Automated risk identification: Automating the risk identification process at scale would benefit device manufacturers as well as the security research community. This would require identifying relevant documents on the web and parsing a huge volume of unstructured documents, while at the same time being able to relate various ML concepts. \halla{
The automated tool also needs to be interfaced with the state-of-the-art vulnerability databases and repositories of peer-reviewed research works so that it can even identify emerging threats in AI and medical devices;
}, \textbf{(2)} Building personalized spatial and temporal risk profiles per patient: Our case study shows that attacks on ML-enabled medical systems cause more damage to certain patients than others. Moreover, a patient is not equally vulnerable at all points of time. An interesting research problem is to study patients' data in more detail to develop customized spatial and temporal risk profiles for every patient; and, \textbf{(3)} Efficient risk mitigation techniques: This involves designing attack-resilient ML models, determining accountable entity and enforcing accountability in risk mitigation, accounting for the costs and deployment scenario.

\vspace{5mm}
\noindent{
\fbox{\parbox{\columnwidth-10pt}{We have made our code and datasets publicly available at: \url{https://gm-repo.github.io/Security-MEDAI/}}}}

\section*{Acknowledgements}
This project is supported by collaborative research funding from the National Research Council of Canada’s Digital Health and Geospatial Analytics Program, UBC Four Year Fellowships (FYF), and the Natural Sciences and Engineering Research Council of Canada (NSERC).

\bibliographystyle{IEEETrans}
\bibliography{sections/cite}   

\begin{thebibliography}{10}

\bibitem{fdaml}
{U.S. FDA}.
\newblock {Artificial Intelligence and Machine Learning (AI/ML)-Enabled Medical Devices}.
\newblock {URL: \url{https://www.fda.gov/medical-devices/software-medical-device-samd/artificial-intelligence-and-machine-learning-aiml-enabled-medical-devices}, Last accessed: Nov 30, 2023}.

\bibitem{chen2020ecgadv}
Huangxun Chen, Chenyu Huang, Qianyi Huang, Qian Zhang, and Wei Wang.
\newblock {Ecgadv: Generating adversarial electrocardiogram to misguide arrhythmia classification system}.
\newblock In {\em AAAI Conference on Artificial Intelligence}, volume~34, pages 3446--3453, 2020.

\bibitem{albattah2023detection}
Albatul Albattah and Murad~A Rassam.
\newblock {Detection of Adversarial Attacks against the Hybrid Convolutional Long Short-Term Memory Deep Learning Technique for Healthcare Monitoring Applications}.
\newblock {\em Applied Sciences}, 13(11):6807, 2023.

\bibitem{lal2021adversarial}
Sheeba Lal, Saeed~Ur Rehman, Jamal~Hussain Shah, Talha Meraj, Hafiz~Tayyab Rauf, Robertas Dama{\v{s}}evi{\v{c}}ius, Mazin~Abed Mohammed, and Karrar~Hameed Abdulkareem.
\newblock {Adversarial attack and defence through adversarial training and feature fusion for diabetic retinopathy recognition}.
\newblock {\em Sensors}, 21(11):3922, 2021.

\bibitem{9313421}
Amiel Meiseles, Ishai Rosenberg, Yair Motro, Lior Rokach, and Jacob Moran-Gilad.
\newblock {Adversarial Vulnerability of Deep Learning Models in Analyzing Next Generation Sequencing Data}.
\newblock In {\em IEEE BIBM}, pages 464--468, 2020.

\bibitem{joel2021adversarial}
Marina~Z Joel, Sachin Umrao, Enoch Chang, Rachel Choi, Daniel Yang, James Duncan, Antonio Omuro, Roy Herbst, Harlan~M Krumholz, Sanjay Aneja, et~al.
\newblock {Adversarial attack vulnerability of deep learning models for oncologic images}.
\newblock {\em MedRxiv}, 2021.

\bibitem{bortsova2021adversarial}
Gerda Bortsova, Cristina Gonz{\'a}lez-Gonzalo, Suzanne~C Wetstein, Florian Dubost, Ioannis Katramados, Laurens Hogeweg, Bart Liefers, Bram van Ginneken, Josien~PW Pluim, Mitko Veta, et~al.
\newblock {Adversarial attack vulnerability of medical image analysis systems: Unexplored factors}.
\newblock {\em Medical Image Analysis}, 73:102141, 2021.

\bibitem{menon2021covid}
Karthika Menon, V~Khushi Bohra, Lakshana Murugan, Kavya Jaganathan, and Chamundeswari Arumugam.
\newblock {COVID-19 Diagnosis from Chest X-Ray Images Using Convolutional Neural Networks and Effects of Data Poisoning}.
\newblock In {\em ICCSA}, pages 508--521, 2021.

\bibitem{vargas2020understanding}
Danilo~Vasconcellos Vargas and Jiawei Su.
\newblock {Understanding the one-pixel attack: Propagation maps and locality analysis}.
\newblock In {\em CEUR Workshop Proceedings}, volume 2640, 2020.

\bibitem{levy2022personalized}
Tamar Levy-Loboda, Eitam Sheetrit, Idit~F Liberty, Alon Haim, and Nir Nissim.
\newblock {Personalized insulin dose manipulation attack and its detection using interval-based temporal patterns and machine learning algorithms}.
\newblock {\em {Journal of Biomedical Informatics}}, 132:104129, 2022.

\bibitem{chen2022adversarial}
Yuzhong Chen, Jiadong Yan, Mingxin Jiang, Tuo Zhang, Zhongbo Zhao, Weihua Zhao, Jian Zheng, Dezhong Yao, Rong Zhang, Keith~M Kendrick, et~al.
\newblock {Adversarial learning based node-edge graph attention networks for autism spectrum disorder identification}.
\newblock {\em IEEE Transactions on Neural Networks and Learning Systems}, 2022.

\bibitem{yu2023perturbing}
Jianfeng Yu, Kai Qiu, Pengju Wang, Caixia Su, Yufeng Fan, and Yongfeng Cao.
\newblock {Perturbing BEAMs: EEG adversarial attack to deep learning models for epilepsy diagnosing}.
\newblock {\em BMC Medical Informatics and Decision Making}, 23(1):115, 2023.

\bibitem{nielsen2022investigating}
Christopher Nielsen, Anup Tuladhar, and Nils~D Forkert.
\newblock {Investigating the Vulnerability of Federated Learning-Based Diabetic Retinopathy Grade Classification to Gradient Inversion Attacks}.
\newblock In {\em International Workshop on Ophthalmic Medical Image Analysis}, pages 183--192, 2022.

\bibitem{mangaokar2020jekyll}
Neal Mangaokar, Jiameng Pu, Parantapa Bhattacharya, Chandan~K Reddy, and Bimal Viswanath.
\newblock {Jekyll: Attacking medical image diagnostics using deep generative models}.
\newblock In {\em IEEE EuroS\&P}, pages 139--157, 2020.

\bibitem{hu2022adversarial}
Lei Hu, Da-Wei Zhou, Xiang-Yu Guo, Wen-Hao Xu, Li-Ming Wei, and Jun-Gong Zhao.
\newblock {Adversarial training for prostate cancer classification using magnetic resonance imaging}.
\newblock {\em Quantitative Imaging in Medicine and Surgery}, 12(6):3276, 2022.

\bibitem{ma2021understanding}
Xingjun Ma, Yuhao Niu, Lin Gu, Yisen Wang, Yitian Zhao, James Bailey, and Feng Lu.
\newblock {Understanding adversarial attacks on deep learning based medical image analysis systems}.
\newblock {\em Pattern Recognition}, 110:107332, 2021.

\bibitem{dreamed}
{U.S. FDA}.
\newblock {DreaMed Advisor Pro}.
\newblock {URL: \url{https://www.accessdata.fda.gov/scripts/cdrh/cfdocs/cfpmn/denovo.cfm?id=DEN170043}, Last accessed: Nov 30, 2023}.

\bibitem{misc_diabetes_34}
Michael Kahn.
\newblock {Diabetes}.
\newblock UCI Machine Learning Repository.
\newblock {DOI}: https://doi.org/10.24432/C5T59G.

\bibitem{Mouri_Badireddy_2023}
MIchelle Mouri and Madhu Badireddy.
\newblock {Hyperglycemia}.
\newblock {URL: \url{https://www.ncbi.nlm.nih.gov/books/NBK430900/}, Last accessed: Nov 30, 2023}.

\bibitem{rubin2020deep}
Harry Rubin-Falcone, Ian Fox, and Jenna Wiens.
\newblock {Deep Residual Time-Series Forecasting: Application to Blood Glucose Prediction.}
\newblock In {\em {KDH@ ECAI}}, pages 105--109, 2020.

\bibitem{rasmussen2022blurtooth}
Kasper Rasmussen.
\newblock {BLURtooth: Exploiting Cross- Transport Key Derivation in Bluetooth Classic and Bluetooth Low Energy}.
\newblock In {\em {AsiaCCS}}, 2022.

\bibitem{Dreamedperipherals}
{Dreamed Diabetes Ltd.}
\newblock {DreaMed Advisor Pro: Manual For Personal Use iOS}.
\newblock {URL: \url{https://dreamed-diabetes.com/wp-content/uploads/2019/06/Dreamed-Advisor-Pro-iOS-Patient-IFU.pdf}, Last accessed: Nov 30, 2023}.

\bibitem{dupont2020matter}
Guillaume Dupont, Daniel~Ricardo dos Santos, Elisa Costante, Jerry Den~Hartog, and Sandro Etalle.
\newblock {A matter of life and death: analyzing the security of healthcare networks}.
\newblock In {\em {SEC 2020: ICT Systems Security and Privacy Protection}}, pages 355--369, 2020.

\bibitem{mortazavi2014electromagnetic}
SMJ Mortazavi, M~Gholampour, M~Haghani, G~Mortazavi, and AR~Mortazavi.
\newblock {Electromagnetic radiofrequency radiation emitted from GSM mobile phones decreases the accuracy of home blood glucose monitors}.
\newblock {\em {Journal of Biomedical Physics \& Engineering}}, 4(3):111, 2014.

\bibitem{hulsen2019big}
Tim Hulsen, Saumya~S Jamuar, Alan~R Moody, Jason~H Karnes, Orsolya Varga, Stine Hedensted, Roberto Spreafico, David~A Hafler, and Eoin~F McKinney.
\newblock {From big data to precision medicine}.
\newblock {\em {Frontiers in medicine}}, 6:34, 2019.

\bibitem{georgakopoulos2016weakly}
Spiros~V Georgakopoulos, Dimitris~K Iakovidis, Michael Vasilakakis, Vassilis~P Plagianakos, and Anastasios Koulaouzidis.
\newblock {Weakly-supervised convolutional learning for detection of inflammatory gastrointestinal lesions}.
\newblock In {\em IEEE IST}, pages 510--514, 2016.

\bibitem{issa2007personalized}
Amalia~M Issa.
\newblock {Personalized medicine and the practice of medicine in the 21st century}.
\newblock {\em McGill Journal of Medicine: MJM}, 10(1):53, 2007.

\bibitem{amin2022two}
Muhammad~Shahid Amin, Jamal~Hussain Shah, Mussarat Yasmin, Ghulam~Jillani Ansari, Muhamamd~Attique Khan, Usman Tariq, Ye~Jin Kim, and Byoungchol Chang.
\newblock {A two-stream fusion assisted deep learning framework for stomach diseases classification}.
\newblock {\em CMC-Comput. Mater. Contin}, 73:4423--4439, 2022.

\bibitem{9251568}
Michal Byra, Grzegorz Styczynski, Cezary Szmigielski, Piotr Kalinowski, Lukasz Michalowski, Rafal Paluszkiewicz, Bogna Ziarkiewicz-Wroblewska, Krzysztof Zieniewicz, and Andrzej Nowicki.
\newblock {Adversarial attacks on deep learning models for fatty liver disease classification by modification of ultrasound image reconstruction method}.
\newblock In {\em IEEE IUS}, pages 1--4, 2020.

\bibitem{zhou2022robustness}
Xugui Zhou, Maxfield Kouzel, and Homa Alemzadeh.
\newblock {Robustness testing of data and knowledge driven anomaly detection in cyber-physical systems}.
\newblock In {\em IEEE/IFIP DSN-W}, pages 44--51, 2022.

\bibitem{nguyen2022novel}
TV~Nguyen, MA~Dakka, SM~Diakiw, MD~VerMilyea, M~Perugini, JMM Hall, and D~Perugini.
\newblock {A novel decentralized federated learning approach to train on globally distributed, poor quality, and protected private medical data}.
\newblock {\em Scientific Reports}, 12(1):8888, 2022.

\bibitem{paul2020mitigating}
Rahul Paul, Matthew Schabath, Robert Gillies, Lawrence Hall, and Dmitry Goldgof.
\newblock {Mitigating adversarial attacks on medical image understanding systems}.
\newblock In {\em IEEE ISBI}, pages 1517--1521, 2020.

\bibitem{mitre-cve}
{Mitre}.
\newblock {Common Vulnerabilities and Exposures}.
\newblock {URL: \url{https://cve.mitre.org/}, Last accessed: Nov 30, 2023}.

\bibitem{nist-cve}
{Information Technology Laboratory, USA}.
\newblock {National Vulnerability Database}.
\newblock {URL: \url{https://nvd.nist.gov/vuln/search}, Last accessed: Nov 30, 2023}.

\bibitem{ecgvul1}
{Mitre}.
\newblock {Conexus Telemetry Protocol vulnerability}.
\newblock {URL: \url{https://cve.mitre.org/cgi-bin/cvename.cgi?name=CVE-2019-6538}, Last accessed: Nov 30, 2023}.

\bibitem{wang2021can}
Wei Wang, Yao Yao, Xin Liu, Xiang Li, Pei Hao, and Ting Zhu.
\newblock {I can see the light: Attacks on autonomous vehicles using invisible lights}.
\newblock In {\em ACM SIGSAC CCS}, pages 1930--1944, 2021.

\bibitem{ir1}
{Mitre}.
\newblock {Sony IPELA E Series Camera vulnerability (1)}.
\newblock {URL: \url{https://cve.mitre.org/cgi-bin/cvename.cgi?name=CVE-2018-3938}, Last accessed: Nov 30, 2023}.

\bibitem{window2}
{Mitre}.
\newblock {Windows 7 vulnerability (2)}.
\newblock {URL: \url{https://cve.mitre.org/cgi-bin/cvename.cgi?name=CVE-2019-5921}, Last accessed: Nov 30, 2023}.

\bibitem{fdapremarket}
{U.S. FDA}.
\newblock {Content of Premarket Submissions for Management of Cybersecurity in Medical Devices}.
\newblock {URL: \url{https://www.fda.gov/media/86174/download}, Last accessed: Nov 30, 2023}.

\bibitem{endo-1}
{Mitre}.
\newblock {Shekar Endoscope vulnerability (1)}.
\newblock {URL: \url{https://cve.mitre.org/cgi-bin/cvename.cgi?name=CVE-2017-10722}, Last accessed: Nov 30, 2023}.

\bibitem{xray-1}
{Mitre}.
\newblock {GE Healthcare Discovery vulnerability}.
\newblock {URL: \url{https://cve.mitre.org/cgi-bin/cvename.cgi?name=CVE-2014-7232}, Last accessed: Nov 30, 2023}.

\bibitem{aami2007ansi}
AAMI.
\newblock {ANSI/AAMI/ISO 14971: 2007/(R) 2010, Medical devices—Application of risk management to medical devices}.

\bibitem{international2015iec}
International~Electrotechnical Commission et~al.
\newblock {IEC 62304: 2006/A1: 2015}.
\newblock {\em Medical device software-Software life-cycle processes}, 2015.

\bibitem{webcam}
{Mitre}.
\newblock {H264WebCam vulnerability}.
\newblock {URL: \url{https://cve.mitre.org/cgi-bin/cvename.cgi?name=CVE-2010-2349}, Last accessed: Nov 30, 2023}.

\bibitem{idx}
{U.S. FDA}.
\newblock {IDx-DR v2.3}.
\newblock {URL: \url{https://www.accessdata.fda.gov/scripts/cdrh/cfdocs/cfpmn/pmn.cfm?ID=K213037}, Last accessed: Nov 30, 2023}.

\bibitem{annas2003hipaa}
George~J Annas.
\newblock {HIPAA regulations: a new era of medical-record privacy?}
\newblock {\em New England Journal of Medicine}, 348:1486, 2003.

\bibitem{phillips1}
{Mitre}.
\newblock {Philips MRI 1.5T and MRI 3T vulnerability (1)}.
\newblock {URL: \url{https://cve.mitre.org/cgi-bin/cvename.cgi?name=CVE-2021-26262}, Last accessed: Nov 30, 2023}.

\bibitem{suarez2015compartmentation}
Guillermo Suarez-Tangil, Juan~E Tapiador, and Pedro Peris-Lopez.
\newblock {Compartmentation policies for android apps: A combinatorial optimization approach}.
\newblock In {\em NSS 2015}, pages 63--77, 2015.

\bibitem{eykholturet}
Kevin Eykholt, Taesung Lee, Douglas Schales, Jiyong Jang, and Ian Molloy.
\newblock {URET: Universal Robustness Evaluation Toolkit (for Evasion)}.
\newblock In {\em {USENIX Security 23}}, pages 3817--3833, 2023.

\bibitem{dnav}
{U.S. FDA}.
\newblock {d-Nav System}.
\newblock {URL: \url{https://www.accessdata.fda.gov/scripts/cdrh/cfdocs/cfpmn/pmn.cfm?ID=K181916}, Last accessed: Nov 30, 2023}.

\bibitem{marling2020ohiot1dm}
Cindy Marling and Razvan Bunescu.
\newblock {The OhioT1DM dataset for blood glucose level prediction: Update 2020}.
\newblock In {\em CEUR workshop proceedings}, volume 2675, page~71. NIH Public Access, 2020.

\bibitem{biggio2013evasion}
Battista Biggio, Igino Corona, Davide Maiorca, Blaine Nelson, Nedim {\v{S}}rndi{\'c}, Pavel Laskov, Giorgio Giacinto, and Fabio Roli.
\newblock {Evasion attacks against machine learning at test time}.
\newblock In {\em ECML PKDD 2013, Proceedings, Part III 13}, pages 387--402, 2013.

\bibitem{finlayson2019adversarial}
Samuel~G Finlayson, John~D Bowers, Joichi Ito, Jonathan~L Zittrain, Andrew~L Beam, and Isaac~S Kohane.
\newblock {Adversarial attacks on medical machine learning}.
\newblock {\em Science}, 363(6433):1287--1289, 2019.

\bibitem{finlayson2018adversarial}
Samuel~G Finlayson, Hyung~Won Chung, Isaac~S Kohane, and Andrew~L Beam.
\newblock {Adversarial attacks against medical deep learning systems}.
\newblock {\em arXiv preprint arXiv:1804.05296}, 2018.

\bibitem{carlini2017towards}
Nicholas Carlini and David Wagner.
\newblock {Towards evaluating the robustness of neural networks}.
\newblock In {\em {IEEE SP}}, pages 39--57, 2017.

\bibitem{Hern_2017}
Alex Hern.
\newblock {Hacking risk leads to recall of 500,000 pacemakers due to patient death fears}.
\newblock {URL: \url{https://www.theguardian.com/technology/2017/aug/31/hacking-risk-recall-pacemakers-patient-death-fears-fda-firmware-update}, Last accessed: Nov 30, 2023}.

\bibitem{Lexology_2023}
Leigh Day.
\newblock {Lawyers investigate claims following further recall of `Assurity' and `Endurity' Abbott Laboratories pacemakers}.
\newblock {URL: \url{https://www.lexology.com/library/detail.aspx?g=08a4668c-cea0-4f60-b348-394ae5209ebc}, Last accessed: Nov 30, 2023}.

\bibitem{Gupta_2013}
Sanjay Gupta.
\newblock {Dick Cheney's heart}.
\newblock {URL: \url{https://www.cbsnews.com/news/dick-cheneys-heart/}, Last accessed: Nov 30, 2023}.

\bibitem{meier2003improving}
J.D. Meier and Microsoft Corporation.
\newblock {\em {Improving Web Application Security: Threats and Countermeasures}}.
\newblock Patterns \& practices. Microsoft press, 2003.

\bibitem{liu2020failure}
Hu-Chen Liu, Li-Jun Zhang, Ye-Jia Ping, and Liang Wang.
\newblock {Failure mode and effects analysis for proactive healthcare risk evaluation: a systematic literature review}.
\newblock {\em Journal of evaluation in clinical practice}, 26(4):1320--1337, 2020.

\bibitem{mahler2020new}
Tom Mahler, Yuval Elovici, and Yuval Shahar.
\newblock {A new methodology for information security risk assessment for medical devices and its evaluation}.
\newblock {\em arXiv preprint arXiv:2002.06938}, 2020.

\bibitem{yaqoob2019integrated}
Tahreem Yaqoob, Haider Abbas, and Narmeen Shafqat.
\newblock {Integrated security, safety, and privacy risk assessment framework for medical devices}.
\newblock {\em IEEE journal of biomedical and health informatics}, 24(6):1752--1761, 2019.

\bibitem{abraham2015common}
R~Abraham, D~Arora, M~Coles, M~Eckert, M~Heitman, A~Manion, S~Moore, S~Romanowsky, K~Scarfone, J~Stuppi, et~al.
\newblock {Common vulnerability scoring system v3. 0: Specification document}.
\newblock {\em First}, 2015.

\bibitem{spring2021time}
Jonathan Spring, Eric Hatleback, Allen Householder, Art Manion, and Deana Shick.
\newblock {Time to Change the CVSS?}
\newblock {\em IEEE S\&P}, 19(2):74--78, 2021.

\bibitem{wu2021medical}
Eric Wu, Kevin Wu, Roxana Daneshjou, David Ouyang, Daniel~E Ho, and James Zou.
\newblock {How medical AI devices are evaluated: limitations and recommendations from an analysis of FDA approvals}.
\newblock {\em Nature Medicine}, 27(4):582--584, 2021.

\end{thebibliography}

\end{document}